\documentclass[aps,prb,10pt,twocolumn,superscriptaddress]{revtex4}

\usepackage{graphicx}
\usepackage{amssymb, amsmath}
\usepackage{color}
\usepackage{ulem}
\usepackage{bm}
\usepackage{graphicx}
\usepackage{subfigure}
\usepackage{dcolumn}
\def\I{\uppercase\expandafter{\romannumeral 1}}
\def\II{\uppercase\expandafter{\romannumeral 2}}
\def\III{{\uppercase\expandafter{\romannumeral 3}}}
\def\IV{{\uppercase\expandafter{\romannumeral 4}}}
\def\V{{\uppercase\expandafter{\romannumeral 5}}}
\def\VI{{\uppercase\expandafter{\romannumeral 6}}}
\def\VII{{\uppercase\expandafter{\romannumeral 7}}}
\def\i{\lowercase\expandafter{\romannumeral 1}}
\def\ii{\lowercase\expandafter{\romannumeral 2}}
\def\iii{{\lowercase\expandafter{\romannumeral 3}}}
\def\iv{{\lowercase\expandafter{\romannumeral 4}}}
\def\v{{\lowercase\expandafter{\romannumeral 5}}}
\def\vi{{\lowercase\expandafter{\romannumeral 6}}}
\def\vii{{\lowercase\expandafter{\romannumeral 7}}}

\def\angstrom{\mbox{\normalfont\AA}}

\def\nn{\nonumber\\}

\def\angstrom{\mbox{\normalfont\AA}}

\def\nn{\nonumber\\}

\begin{document}

\title{The pseudo-Landau-level representation of twisted bilayer graphene: band topology and the implications on the correlated insulating phase}

\author{Jianpeng Liu}
\affiliation{Department of Physics, Hong Kong University of Science and Technology, Kowloon, Hong Kong}

\author{Junwei Liu}
\affiliation{Department of Physics, Hong Kong University of Science and Technology, Kowloon, Hong Kong}

\author{Xi Dai}
\affiliation{Department of Physics, Hong Kong University of Science and Technology, Kowloon, Hong Kong}

\begin{abstract}
We propose that the electronic structure of twisted bilayer graphene (TBG) can be understood as Dirac fermions coupled with opposite pseudo magnetic fields generated by the moir\'e pattern. The two low-energy flat bands from each monolayer valley originate from the two zeroth pseudo Landau levels of Dirac fermions under such opposite effective magnetic fields, which have opposite sublattice polarizations and carry opposite Chern numbers $\pm1$, giving rise to helical edge states in the gaps below and above the low-energy bulk bands near the first magic angle.  We argue that small Coulomb interactions would split the eight-fold degeneracy (including valley and physical spin) of these zeroth pseudo Landau levels, and may lead to  insulating phases with non-vanishing Chern numbers at integer fillings. 
Besides, we show that all the high-energy bands below or above the flat bands are also topologically nontrivial in the sense that for each valley the sum of their Berry phases is quantized as $\pm\pi$. Such quantized Berry phases give rise to nearly flat edge states, which are dependent on truncations on the moir\'e length scale. Our work provides a complete and clear picture for the electronic structure and topological properties of TBG, and has significant implications on the natrue of the correlated insulating phase observed in experiments.
\end{abstract}

\maketitle

Twisted bilayer graphene (TBG) is an engineered system with one graphene layer stacked on top of the other and rotated by a twisted angle $\theta$, which exhibits various interesting properties \cite{santos-tbg-prl07, macdonald-pnas11, li-vhs-11, smet-qhe-tbg-prl11,qhe-tbg-prl12,he-prl11,tbg-nonabelian-prl12}. Around the so called ``magic angles" the low-energy electronic structures of TBG are characterized by four nearly flat bands contributed by the two monolayer valleys \cite{macdonald-pnas11}, and these flat low-energy bands are believed to be responsible for the correlated insulating phases \cite{cao-nature18-mott, sharpe-tbg-19,choi-tbg-stm,kerelsky-tbg-stm} and unusual superconductivity \cite{cao-nature18-supercond}.
Numerous theoretical attempts have been made to understand the electronic structures \cite{po-prx18, yuan-prb18, koshino-prx18, kang-prx18, song-tbg-18, po-tbg2, hejazi-arxiv18,origin-magic-angle-tbg18,ramires-prl18, pal-kindermann-arxiv18, ll-tbg-lian}, the structural properties \cite{koshino-tbg-prb17,jain-tbg-structure,tbg-D6-arxiv18}, the correlated insulating phase \cite{po-prx18,sboychakov-arxiv-18, isobe-prx18, xu-lee-prb18, huang-arxiv-18,liu-prl18,rademaker-prb18, venderbos-prb18, kang-tbg-correlation-arxiv18, xie-tbg-2018, zaletel-tbg-2019}, and the mechanism of superconductivity \cite{xu-prl18,po-prx18,isobe-prx18,wu-prl18,wu-xu-arxiv-18,lian-arxiv-18, huang-arxiv-18,liu-prl18, venderbos-prb18, nematic-tbg-arxiv18, wu-tbg-chiral-arxiv18}. However, up to now, the nature of the correlated insulating phase and the superconductivity are still obscure. 

Besides many-body effects, the four low-energy bands already exhibits interesting or even puzzling properties at the single-particle level \cite{po-prx18, koshino-prx18, po-tbg2, song-tbg-18,yuan-prb18,kang-prx18}.
In particular recently it has been shown that the four low-energy bands are topologically nontrivial in the sense that they are characterized by odd windings of Wilson loops.
However, despite the numerical evidence \cite{song-tbg-18} and the mathematical classifications \cite{song-tbg-18,yang-tbg18}, how to physically understand and describe the topological nature of the flat bands in TBG is still an open question. On the other hands, so far the topological properties of the high-energy bands have been rarely discussed. To fully understand the unusual electronic and topological properties of TBG, a clear and complete physical picture is needed.

We address these issues in this work and reach the following conclusions. We find that  in the small twist angle limit, the low energy electronic structures of TBG can be viewed as 2D Dirac models under pseudo magnetic fields generated by the moir\'e pattern. The nontrivial topology of the two low-energy bands for each valley originates from the two zeroth pseudo Landau levels (LLs) of Dirac fermions  with such opposite effective magnetic fields. The two zeroth LLs (for each valley) carry  opposite Chern numbers $\pm$1 and possess opposite sublattice polarizations. They are decoupled from each other as a result of an emergent chiral symmetry in the low-energy subspace. This leads to two pairs of helical edge states in the energy gaps below and above the low-energy bands of TBG.
As the four low-energy bands (of the two valleys) in TBG are equivalent to four zeroth pseudo LLs, small Coulomb interactions are expected to split the pseudo LL degeneracy at integer fillings, and could lead to insulating states with fully polarized zeroth pseudo LLs and non-vanishing total Chern numbers, as possibly suggested by the recent experiment on 7/8 filled    TBG around the first magic angle   \cite{sharpe-tbg-19,comment_filling_tbg}. 

In addition to the low-energy bands, we show that for each valley the high-energy bands below and above the low-energy bands are also topologically nontrivial with quantized Berry phases $\pm\pi$. Such quantized Berry phases give rise to two nearly flat edge states in the gaps between the high-energy bands and the low-energy bands, which are dependent on the truncations on the moir\'e length scale. 
Last, we find  that the topological gaps between the high-energy bands and the low-energy bands can be significantly enhanced by atomic corrugations, and that changing the corrugation strength may further drive transitions between  insulating and semimetallic phases.

The paper is organized as follows. In Sec.~\ref{sec:tbg-continuum} we discuss the lattice structure of TBG and introduce the continuum model describing the electronic structures of TBG. In Sec.~\ref{sec:landau} we introduce the pseudo-Landau-level representation of TBG, which provides a clear physical picture for the topological properties of the low-energy bands, and has significant implications on the nature of the correlated insulating phases observed in experiments. In Sec.~\ref{sec:high-energy} we discuss the topological properties of the high-energy bands and the truncation dependence of the edge states. In Sec.~\ref{sec:corrugation}, we discuss in detail the effects of atomic corrugations on the electronic structures of TBG. In Sec.~\ref{sec:summary} we make a summary.

\section{The TBG system and the continuum model}
\label{sec:tbg-continuum}
\begin{figure}
\includegraphics[width=3.0in]{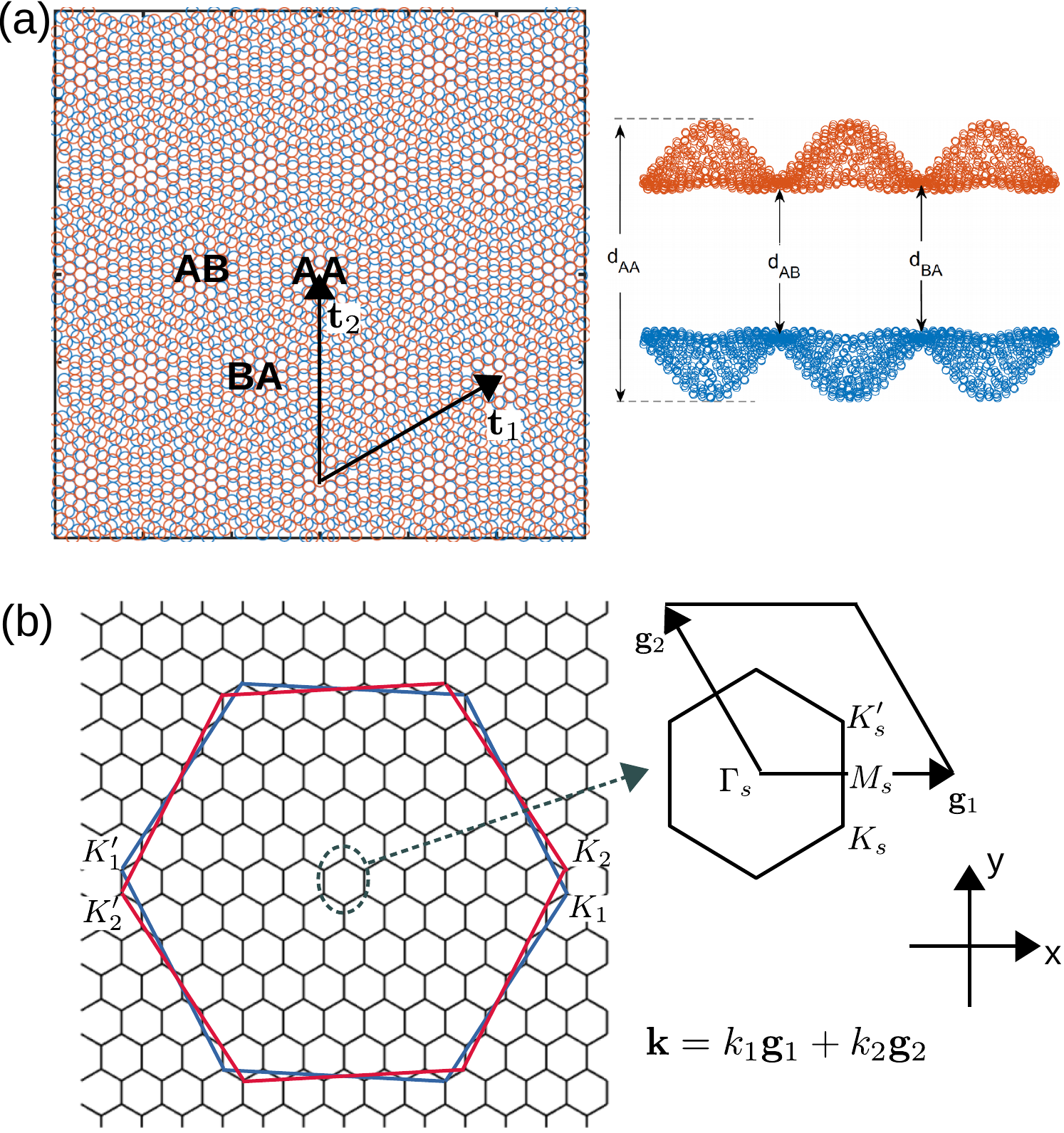}
\caption{(a) Left: a top view of the moir\'e pattern of twisted bilayer graphene for $m\!=\!5$ ($\theta\!\approx\!6.01^{\circ}$). The solid and dashed lines represent lattice truncations through the $AA$ regions and the $AB/BA$ regions respectively. The two arrows denote the lattice vectors. Right: illustration of atomic corrugations. (b) The Brillouin zones of the top monolayer, bottom  monolayer, and the moire\'e supercell are plotted in red, blue, and black lines respectively.}
\label{fig1}
\end{figure}
As shown in Fig.~\ref{fig1}(a), the commensurate moir\'e pattern is formed when the top-layer graphene is rotated with respect to the bottom layer by certain  angles $\{\theta(m)\}$, where $m$ is an integer obeying the condition $\cos{\theta(m)}=(3m^2+3m+1/2)/(3m^2+3m+1)$ \cite{castro-neto-prb12}. The  lattice vectors of the moir\'e superlattice are expressed as $\mathbf{t}_1=(\sqrt{3}L_s/2,L_s/2)$, and $\mathbf{t}_2=(0,L_s)$, where $L_s=\vert\mathbf{t}_1\vert=a/(2\sin{(\theta/2)})$ is  the size of the moir\'e supercell, with $a=2.46\,$\angstrom\ being the lattice constant of monolayer graphene.  The $K$ ($K'$) points of the two monolayers $K_1$ ($K_2'$) and $K_2$ ($K_1'$) are respectively mapped to $K_s$ and $K_s'$ points in the moir\'e supercell Brillouin zone (BZ) as shown in Fig.~\ref{fig1}(b).

Locally homogeneous regions are formed in the moir\'e pattern of TBG. In some regions the $A(B)$ sublattice of the top layer is mostly on top of the same sublattice of the bottom layer, and such regions are dubbed as the ``$AA$" region as shown in Fig.~\ref{fig1}(a); while in some other regions the $B(A)$ sublattice of the top layer is on top of $A(B)$ sublattice of the bottom layer, which are marked as $``AB" (``BA")$ regions. It worth to note that the interlayer distance in TBG varies in real space \cite{uchida-corrugation-prb14}. In the $AB (BA)$ region  the interlayer distance $d_{AB}\!\approx\!3.35\,$\angstrom\, while in the $AA$-stacked region the interlayer distance
$d_{AA}\!\approx\!3.6\,$\angstrom\ \cite{graphite-AA}. Such atomic corrugations  may be modeled as \cite{koshino-prx18}
\begin{equation}
d_z(\mathbf{r})=d_{0}+2d_1\sum_{j=1}^{3}\cos{(\,\mathbf{b}_j\!\cdot\!\mathbf{\delta}(\mathbf{r})\,)}\;,
\label{eq:dz-1}
\end{equation}
where $\mathbf{b}_{1}=(2\pi/a,2\pi/(\sqrt{3}a))$, $\mathbf{b}_2=(-2\pi/a,2\pi/(\sqrt{3}a))$, and $\mathbf{b}_3=\mathbf{b}_1+\mathbf{b}_2$ are three reciprocal lattice vectors of monolayer graphene. $\mathbf{\delta}(\mathbf{r})$ is a 2D vector indicating the local in-plane shift between the carbon atoms in the two layers around position $\mathbf{r}$ in the moir\'e supercell. In the $AA$ region
$\mathbf{\delta}\!\approx\!(0,0)$ while in the AB region $\mathbf{\delta}\!\approx\!(0,a/\sqrt{3})$. We take $d_0=3.433\,$\angstrom\ and $d_1=0.0278\,$\angstrom\ in order to reproduce the interlayer distances in $AA$- and $AB$-stacked bilayer graphene.

The electronic structure of TBG can be described by the Bloch states around the Dirac points  in the two graphene monolayers which mutually tunnel to each other, and can be formulatted by a continuum model proposed by Bistritzer and MacDonald \cite{macdonald-pnas11}. Using such a  continuum model of TBG,  Bistritzer and MacDonald found that  for each monlayer valley  ($K$ or $K'$) there are two low-energy bands the  bandwidths of which vanish recurrently at a series of ``magic angles" starting from $\sim\!1.05^{\circ}$ \cite{macdonald-pnas11}. The states from the two monolayer valleys $K$ and $K'$ are assumed to be decoupled from each other, because the scattering amplitudes are negligbly small at small twist angles \cite{macdonald-pnas11,castro-neto-prb12}.

To be specific, the continuum model describing the TBG system for the $K$ valley  is expressed as
\begin{equation}
H^{+}(\hat{\mathbf{k}})=\begin{pmatrix}
-\hbar v_{F}(\hat{\mathbf{k}}-\mathbf{K}_1)\cdot\mathbf{\sigma} &  Ue^{-i\Delta\mathbf{K}\cdot\mathbf{r}}\\
U^{\dagger}e^{i\Delta\mathbf{K}\cdot\mathbf{r}} & -\hbar v_{F}(\hat{\mathbf{k}}-\mathbf{K}_2)\cdot\mathbf{\sigma}
\end{pmatrix}\;
\label{eq:h+}
\end{equation}
where $v_F$ is the bare Fermi velocity of the Dirac cone in gaphene, $\hat{\mathbf{k}}=-i\partial_\mathbf{r}$, and $\mathbf{K}_1$ and $\mathbf{K}_2$ are the $\mathbf{K}$ points of the bottom and top layers  as shown in Fig.~\ref{fig1}(b). The Pauli matrices $\mathbf{\sigma}=(-\sigma_x,\sigma_y)$ are defined in the space of the $A, B$ sublattices of graphene. The tunneling between the Dirac states in the two layers is described by the $2\times 2$ matrix $U$
\begin{equation}
U=\begin{pmatrix}
u_0 g(\mathbf{r}) & u_0'g(\mathbf{r}-\mathbf{r}_{AB})\\
u_0'g(\mathbf{r}+\mathbf{r}_{AB}) & u_0 g(\mathbf{r})
\end{pmatrix}\;,
\label{eq:u}
\end{equation}
where $\mathbf{r}_{AB}\!=\!(\sqrt{3}L_s/3,0)$,  $u_0'$ and $u_0$ denote the intersublattice and intrasublattice interlayer tunneling amplitudes. $u_0\!<\!u_0'$ if the effects of atomic corrugations are taken into account \cite{koshino-prx18}. $\Delta\mathbf{K}=\mathbf{K}_2-\mathbf{K}_1=(0,4\pi/3L_s)$ is the shift between the Dirac points of the two monolayers, and the phase factor $g(\mathbf{r})$ is defined as
$g(\mathbf{r})=\sum_{j=1}^{3}e^{i\mathbf{q}_j\cdot\mathbf{r}}$,
with $\mathbf{q}_1=(0,4\pi/3L_s)$, $\mathbf{q}_2=(-2\pi/\sqrt{3}L_s,-2\pi/3L_s)$, and $\mathbf{q}_3=(2\pi/\sqrt{3}L_s,-2\pi/3L_s)$.

The continuum model of each valley  has the symmetry generators $C_{3z}$, $C_{2z}\mathcal{T}$,  and $C_{2x}$,
where $\mathcal{T}$ is the time-reversal operation for spinless fermions (i.e., complex conjugation). The two valleys can be mapped to each other by  $\mathcal{T}$, $C_{2z}$, or $C_{2y}$ operations.
Moreover, there is an additional particle-hole-like symmetry which transforms $H^{+}(\hat{\mathbf{k}})$ to
the Hamiltonian of the other valley $H^{-}(-\hat{\mathbf{k}})$:
\begin{equation}
\Lambda\,H^{+}(\hat{\mathbf{k}})\Lambda^{-1}=-H^{-}(-\hat{\mathbf{k}})
\label{eq:ph-symm}
\end{equation}
where $\Lambda=i\tau_y\sigma_x$.

\section{The pseudo-Landau-level representation of twisted bilayer graphene}
\label{sec:landau}

\subsection{The pseudo-Landau-level representation and the band topology}
In this section we show that the two low-energy bands (per valley) in TBG can be represented by the two zeroth pseudo Landau levels carrying opposite Chern numbers. We will focus on the $K$ valley, i.e., the Hamiltonian in Eq.~(\ref{eq:h+}). The Hamiltonian of the other valley can be obtained by a time-reversal operation. 

First we note that the constant wavevectors $\mathbf{K}_1$ and $\mathbf{K}_2$ in Eq.~(\ref{eq:h+}) can be gauged by applying the following transformations to the basis Bloch functions,   
\begin{align}
&\psi_{ls}(\mathbf{r})\to\psi_{ls}(\mathbf{r})e^{i\mathbf{K}_l\cdot\mathbf{r}}\;,
\label{eq:gauge-transform}
\end{align}
where $l=1,2$ and $s=A, B$ refer to the layer and sublattice degrees of freedom respectively.
Then Eq.~(\ref{eq:h+}) becomes
\begin{equation}
H^{+}(\mathbf{k})=
\begin{pmatrix}
-\hbar v_{F}\mathbf{k}\cdot\mathbf{\sigma} &  U\\
U^{\dagger} & -\hbar v_{F}\mathbf{k}\cdot\mathbf{\sigma}
\end{pmatrix}\;.
\label{eq:h+2}
\end{equation}
Next we expand the phase factors $g(\mathbf{r})$ and $\mathbf{g}(\mathbf{r}\pm\mathbf{r}_{AB})$ in Eq.~(\ref{eq:u}) to the linear order of $r/L_s$, and rewrite Eq.~(\ref{eq:h+2}) in the following form
\begin{align}
H^{+}(\mathbf{k})=-\hbar{v}_{F}(\mathbf{k}-\frac{e}{\hbar}\mathbf{A}\tau_y)\cdot\mathbf{\sigma}+3u_0\tau_x.
\label{eq:h+3}
\end{align}
where the Pauli matrices $\mathbf{\tau}$ and $\mathbf{\sigma}$ are defined in the space of the two layers and the two sublattices respectively, with $\mathbf{\sigma}=(-\sigma_x,\sigma_y)$. The effective vector potential $\mathbf{A}\!=\!(2\pi u_0')/(L_s ev_{F})\,(y\, , -x\,)$. In the end we transform to the basis that diagonalizes $\tau_y$, i.e.,
\begin{align}
&\psi_{\alpha,s}(\mathbf{r})=\frac{1}{\sqrt{2}}(\psi_{1,s}(\mathbf{r})+ i\psi_{2,s}(\mathbf{r}))\;,\nn
&\psi_{\beta,s}(\mathbf{r})=\frac{1}{\sqrt{2}}(\psi_{1,s}(\mathbf{r})- i\psi_{2,s}(\mathbf{r}))\;,
\label{eq:basis-transform}
\end{align}
where  $\psi_{ls}(\mathbf{r})$ with layer index $l=1,2$ and sublattice index $s=A, B$ is the Bloch function of the monolayer graphene at $\mathbf{K}_l$ from the $s$ sublattice. In this basis, Eq.~(\ref{eq:h+}) eventually becomes
\begin{equation}
H^{+}(\hat{\mathbf{k}})=\begin{pmatrix}
-\hbar v_{F}(\hat{\mathbf{k}}-\frac{e}{\hbar}\mathbf{A})\cdot\mathbf{\sigma}  & -3i u_0 \\
3iu_0 & -\hbar v_{F}(\hat{\mathbf{k}}+\frac{e}{\hbar}\mathbf{A})\cdot\mathbf{\sigma}
\end{pmatrix}\;.
\label{eq:h+4}
\end{equation}
Again the gauge field $\mathbf{A}\!=\!(2\pi u_0')/(L_s ev_{F})\,(\,y\, , -x\,)$.
Without the off-diagonal term $\pm3iu_0$, Eq.~(\ref{eq:h+4}) is nothing but two Dirac fermions coupled to opposite effective magnetic fields $\pm\mathbf{B}_s=\pm\nabla\times\mathbf{A}$ with the magnitude $B_s\!=\!3u_0'\Delta K/(ev_{F})$, where $\Delta K\!=\!4\pi/(3L_s)$ is the distance between the two Dirac points in the two layers. It is known that $u_0'\!\approx\!0.1\,$eV \cite{macdonald-pnas11,koshino-prx18}, and $\hbar v_F\!\approx\!5.25\,$eV\angstrom, then we estimate $B_s\!\approx\!120$\,T for $\theta\!\approx\!1.08^{\circ}$.

Let us first neglect the off-diagonal term $\pm3iu_0$, then Eq.~(\ref{eq:h+4}) becomes exactly solvable. 
The eigenenergies are just the LLs of the Dirac fermions, $E_{\pm N k}=\pm\hbar\omega_c\sqrt{N}$, with $N\!\ge\!0$ being an integer. The corresponding cyclotron frequency $\omega_c$ and the magnetic length $l_B$ are
\begin{align}
&\hbar\omega_c=\sqrt{\frac{8\pi\hbar v_{F} u_0'}{L_s}}\;,\nn
&l_B=\sqrt{\frac{L_s\hbar v_F}{4\pi u_0'}}\;.
\label{eq:wc-lb}
\end{align}
%
The eigenfunctions of the  upper ($\alpha$) and lower ($\beta$) diagonal blocks of Eq.~(\ref{eq:h+4}) (in the Landau gauge) are expressed as
\begin{align}
&\phi^{\alpha}_{\pm N k} (x,y)=\frac{1}{\sqrt{2L_xl_B}}e^{ikx}
\begin{pmatrix}
\mp\Psi_{N}(\xi)\;\\
\Psi_{N-1}(\xi)
\end{pmatrix}\;\nn
&\phi^{\beta}_{\pm N k} (x,y)=\frac{1}{\sqrt{2L_xl_B}}e^{-ikx}\begin{pmatrix}
\Psi_{N-1}(\xi)\;\\
\pm\Psi_{N}(\xi)
\end{pmatrix}\;,
\label{eq:eigenfunction}
\end{align}
where $\xi\!=\!y/l_B-l_B k$, and $\Psi_N(\xi)\!=\!1/(2^{N/2}\sqrt{N!}\pi^{1/4})e^{-\xi^2/2}H_{N}(\xi)$ is the eigenfunction of the 1D quantum Harmonic oscillator, with $H_{N}(\xi)$ being the Hermite polynominal with $N\!\ge\!0$.  The wavevector $k=2\pi j/L_x$ is the index for the LL degeneracy with  the integer $0\!\le\!j\!\le\!L_x L_y/(2\pi l_B^2)$, and $L_x$ and $L_y$ denote the size of the system along the $x$ and $y$ directions.
These eigenstates have the interesting property that the zeroth LLs in the upper and lower diagonal blocks have exactly opposite sublattice polarizations. The two zeroth pseudo LLs also carry opposite Chern numbers $\pm 1$ \cite{supp_info}, which is the origin of the odd winding pattern of the Wilson loops (see blue lines in Fig.~\ref{fig4}(b)).

Now we   consider the off-diagonal term $\pm 3iu_0$ in Eq.~(\ref{eq:h+4}) (denoted as $H^{+}_{T}$ hereafter) that couples the pseudo LLs in the upper and lower diagonal blocks. First, we note that the coupling term $H^{+}_T$ is \textit{intrasublattice}, but as discussed above the two zeroth pseudo LLs have exactly opposite sublattice polarizations. Therefore the direct coupling within the subspace of zeroth  pseudo LLs exactly vanishes. To be specific, in the pseudo-LL basis $H^{+}_T$ (after transforming to the Landau gauge) can be rewritten as
\begin{align}
&\langle\phi^{\alpha}_{\lambda' N' k'}\vert H^{+}_T\vert\phi^{\beta}_{\lambda N k}\rangle\;\nn
=&\frac{-3iu_0l_B}{2L_x}\Big(\,-\lambda'\Psi_{N'}(l_B k)\Psi_{N-1}(l_B k')\;\nn
&+\lambda\Psi_{N'-1}(l_B k)\Psi_{N}(l_B k')\,\Big)\;,
\label{eq:tunneling-matrix-ll}
\end{align}
where $\lambda, \lambda'=\pm$ denote the upper and lower branches of the Landau levels, and $N\!\ge\!0$. Eq.~(\ref{eq:tunneling-matrix-ll}) clearly indicates that the direct coupling between the two zeroth pseudo LL vanishes, and that the coupling are stronger for higher pseudo LLs with larger $N$ indices. This means that the higher pseudo LLs would be strongly coupled with each other and would lost their topological character.
The  zeroth pseudo LL from the upper (lower) diagonal block could be coupled with the higher LLs from the lower (upper) diagonal block, which would give rise to a finite bandwidth (denoted by $W$) to the otherwise exactly flat zeroth pseudo LLs. However, a straightforward calculation using perturbation theory indicates the leading-order energy correction to the zeroth pseudo LL is on the order of $u_0^3/(\hbar\omega_c)^2$. Therefore, the topological character of the zeroth pseudo LLs is expected to be unchanged as long as the pseudo LL spacing $\hbar\omega_c$ is greater than the bandwidth induced by $u_0$. In Table~\ref{table:llspacing} we show the pseudo LL spacings $\hbar\omega_c$ and the bandwidths $W$ of the low-energy bands at different twist angles in TBG, which is calculated using the continuum model shown in Eq.~(\ref{eq:h+}). Clearly $\hbar\omega_c$ becomes much greater than the low-energy bandwidth $W$ when $m\!\gtrapprox\!15$ ($\theta\!\lessapprox\!2.03^{\circ}$).
\begin{table}[bth]
\caption{Pseudo LL spacings ($\hbar\omega_c$) and the low-energy bandwidths ($W$) at different twist angles (in units of eV)}
\begin{ruledtabular}
\begin{tabular}{lclclclc}
$m$& 15 & 20 & 25 & 30 & 31 & 32
& 33 \\
$\hbar\omega_c$ & 0.441  &  0.384  &  0.344  &  0.315  &  0.310    & 0.305  &  0.300 \\
$W$ & 0.274 &   0.125 &   0.041 &   0.007  &  0.014  &  0.020 &  0.025
\end{tabular}
\end{ruledtabular}
\label{table:llspacing}
\end{table}
%

\subsection{The edge states of the pseudo Landau levels}

\begin{figure}
\includegraphics[width=2.5in]{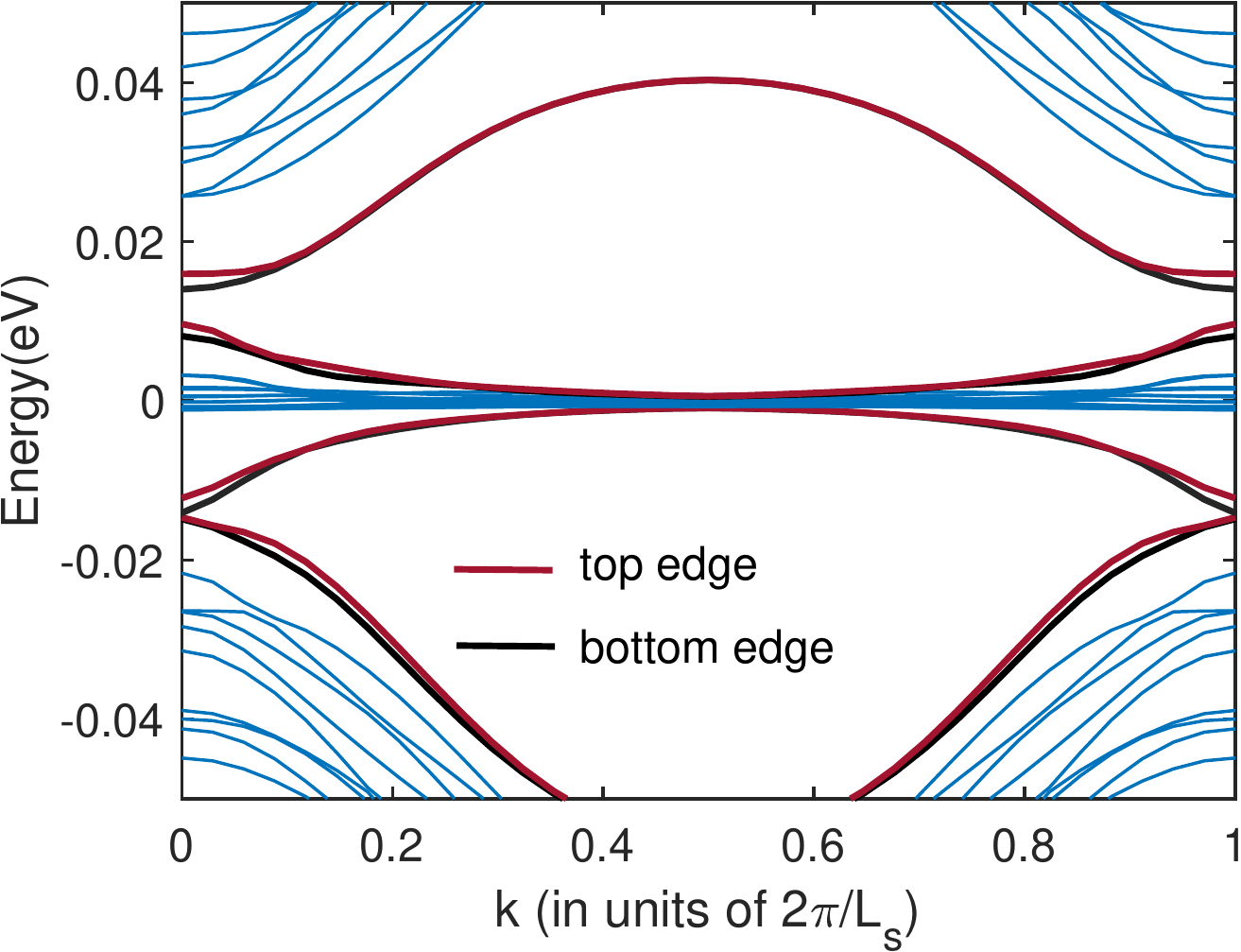}
\caption{The band structure of twisted bilayer graphene at $\theta\!\approx\!1.08^{\circ}$ in a ribbon geometry with the open boundary condition. The red and black lines represent the edge states from the two edges of the ribbon, and blue lines represent the bulk states.}
\label{fig2}
\end{figure}

Neglecting the off-diagonal term $\pm3iu_0$, the two zeroth pseudo LLs  of opposite Chern numbers would give rise to a pair of helical gapless edge states. 
However,  the two zeroth pseudo LLs  could be coupled indirectly by tunneling to the  higher LLs with the amplitude $\sim u_0^3/(\hbar\omega_c)^2$. Thus such high-order couplings will open a gap in the edge states $\sim\!u_0^3/(\hbar\omega_c)^2\!\approx\!5\,$meV  at $\theta\!\approx\!1.08^{\circ}$. On the other hand, the pseudo-LL picture discussed above is valid for the leading-order expansion of $(r/L_s)$. At a non-vanishing twist angle $\theta$, the $\mathcal{O}(r/L_s)^2$ term would also  (weakly) couple the two zeroth pseudo LLs, which would open a gap in the otherwise gapless edge states. Therefore, one expects to see two pairs of slightly gapped helical edge states in the energy gaps below and/or above  the bulk low-energy bands, which are contributed by the two monolayer valleys. 

In Fig.~\ref{fig2} we show the band structure of TBG at the first magic angle $\theta\!\approx\!1.08^{\circ}$, which is calculated using a microscopic Slater-Koster-type tight-binding model \cite{trambly-nanoletter10,moon-tbg-prb13}. To be specific, the hopping integral between two $p_z$ orbitals at different carbon sites $i$ and $j$ (in either of the two layers) is expressed in the Slater-Koster form
\begin{equation}
t(\mathbf{d})=V_{\sigma}\,(\frac{\mathbf{d}\cdot\mathbf{\hat{z}}}{d})^2+V_{\pi}\,(\,1-(\frac{\mathbf{d}\cdot\mathbf{\hat{z}}}{d})^2\,)
\label{eq:hopping}
\end{equation}
where  $V_{\sigma}=V_{\sigma}^{0}\,e^{-(r-d_c)/\delta_0}$, and
$V_{\pi}=V_{\pi}^{0}\,e^{-(r-a_0)/\delta_0}$. $\mathbf{d}=(d_x, d_y, d_z)$ is  the displacement vector between the two carbon sites. $a_0=a/\sqrt{3}=1.42\,$\angstrom, $d_c=3.35\,$\angstrom\ is the interlayer distance in AB-stacked bilayer graphene, and $\delta_0=0.184\,a$. $V_{\sigma}^{0}=0.48\,$eV and $V_{\pi}^{0}=-2.7\,$eV.  The atomic corrugations are modeled by Eq.~(\ref{eq:dz-1}), and their effects can be taken into account in the tight-binding model by plugging Eq.~(\ref{eq:dz-1}) into Eq.~(\ref{eq:hopping}).   

We have constructed a ribbon of TBG at the first magic angle using the above Slater-Koster tight-binding model. The ribbon has translation symmetry along the $y$ direction, and has a finite width $\sim\,$68\,nm (6 moir\'e cells) along the $x$ direction. The red and black lines in Fig.~\ref{fig2} denote the states localized at the two edges, while the blue lines represent the bulk bands. Clearly at each edge there are two pairs of slightly gapped helical edge states in the two bulk energy gaps below and above the low-energy bands. The gaps in the edge states are due to the couplings between the LLs in the two blocks, and the magnitudes of the gaps $\sim\,$3-10\,meV, in agreement with the previous argument. As a comparison, the bulk band structure at $\theta\!\approx\!1.08^{\circ}$ calculated using the same tight-binding model is shown in Fig.~\ref{fig4}(a) in blue lines.

\subsection{Robustness of the Wilson loops at finite twist angles}

\subsubsection{Symmetry analysis of the Wilson-loop operators}
\label{sec：wl-symmetry}
At finite twist angles the $\mathcal{O}(r/L_s)^2$ terms become non-negligible, which would directly couple the two zeroth pseudo LLs. However, numerically the Wilson loops of the flat bands in TBG retain the their topological character even at  large twist angles \cite{supp_info}. It turns out that the odd winding pattern of the Wilson loops (see the blue circles in Fig.~\ref{fig4}(b)) is protected $C_{2z}\mathcal{T}$  \cite{song-tbg-18,yang-tbg18} and $C_{2x}$ symmetries \cite{song-tbg-18}. 

To be explicit, denoting the Wilson loop of the two flat bands at $k_2$ (integrated along $k_1$) by a $2\times 2$ matrix $\hat{w}(k_2)$, we find that they obey the following relationship due to the constriants from the $C_{2z}\mathcal{T}$ and $C_{2x}$ symmetries
\begin{align}
&w_{mn}(k_2)=-\xi_{m}\xi_nw^{*}_{mn}(k_2)-j_n\delta_{mn}\;,\nn
&w_{mn}(k_2)=w_{mn}(-k_2)- j'_n\delta_{mn}\;,\nn
\label{eq:wl-constriants}
\end{align}
where $w_{mn}(k_2)$ is the matrix element of $\hat{w}(k_2)$, $j_n$ and $j'_n$ are arbitrary integers, and $m$, $n$ are the band indices. $\xi_m$, $\xi_n=\pm 1$ are the eigenvalues of the $C_{2z}\mathcal{T}$ operator for the energy bands $m$ and $n$.
The first line of Eq.~(\ref{eq:wl-constriants}) is from the $C_{2z}\mathcal{T}$ symmetry, which indicates that the diagonal element of the Wilson loop operator $w_{nn}(k_2)\!=\!j_n$. Moreover, for $\xi_m\xi_n=\pm 1$ the off-diagonal element $w_{mn}(k_2)=\mp w_{mn}^{*}(k_2)$ for $m\neq n$. Therefore, for a 2-band system there is only one parameter describing the variation of $\hat{w}(k_2)$ with respect to $k_2$
\begin{equation}
\hat{w}(k_2)=\begin{pmatrix}
j_1 & 0 \\
0   & j_2
\end{pmatrix} + \begin{cases}
 d_y(k_2)\sigma_y,  \textrm{  if } \xi_1\xi_2=1 \\
d_x(k_2)\sigma_x,    \textrm{  if } \xi_1\xi_2=-1
\end{cases}\;,
\label{eq:wl-c2t}
\end{equation}
where $j_1$ and $j_2$ are two arbitrary integers and $\sigma_x$ and $\sigma_y$ are the Pauli matrices in the space of the two bands.
Eq.~(\ref{eq:wl-c2t}) guarantees that a  band touching point in the Wilson-loop spectrum is topologically stable, which cannot be gapped out unless two touching points meet each other and get annihilated. This is consistent with the conclusion in Ref.~\onlinecite{song-tbg-18} and ~\onlinecite{yang-tbg18}. Moreover,  Eq.~(\ref{eq:wl-c2t}) also suggests that the sum of the Wilson-loop eigenvalues of the two flat bands can only take integer values. The second line of Eq.~(\ref{eq:wl-constriants}) is from the $C_{2x}$ symmetry, which dictates that the Wilson loop at $k_2$ equals to that at $-k_2$ \cite{song-tbg-18} with some integer ambiguity in the trace. Such a constriant requires that the degeneracy points in the Wilson loop have to either occur in pairs at $k_2$ and $-k_2$ or at the high-symmetry points $k_2\!=\!0$ and/or $0.5$ (in reduced coordinates). This is why the Wilson-loop spectra retain their topological character even at finite twist angles.


\subsubsection{Robustness of the Wilson loops against microscopic perturbations}
\label{appen:wilson-loop-microscopic}
\begin{figure}[hbt]
\includegraphics[width=3.4in]{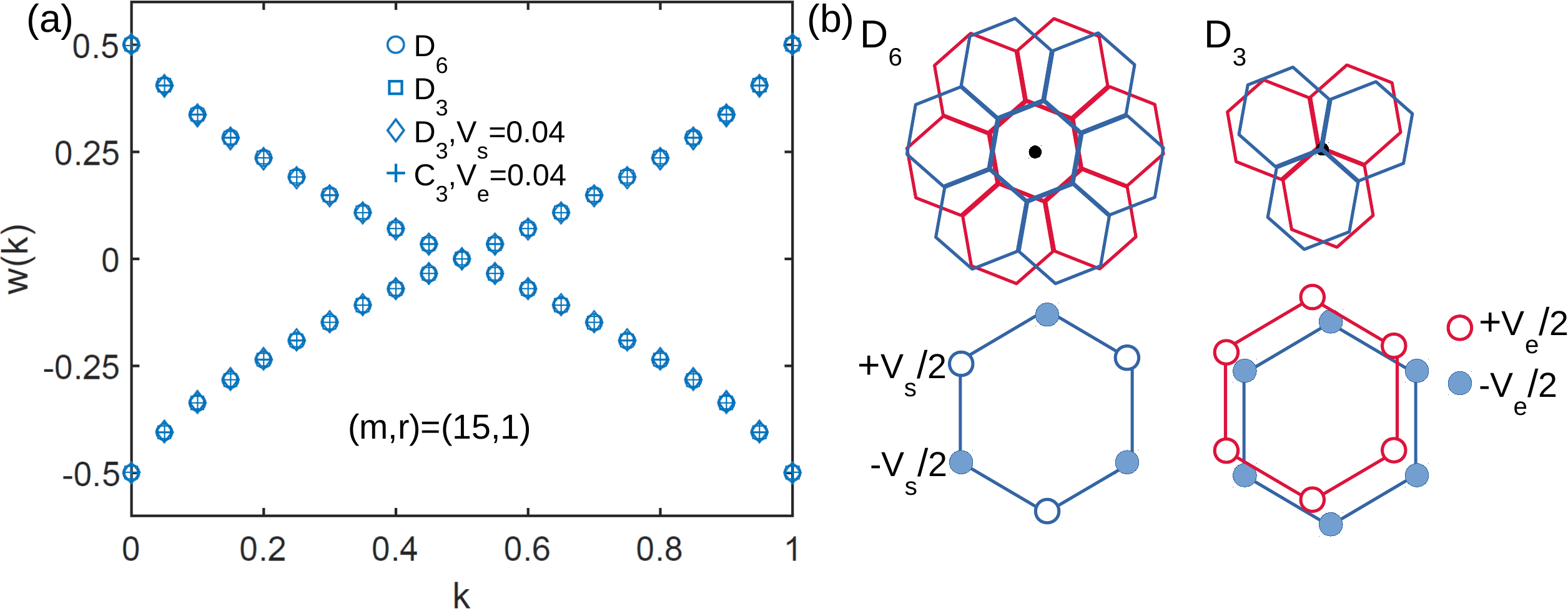}
\caption{(a) The Wilson loops of the four low-energy bands of twisted bilayer graphene at $m\!=\!15$ ($\theta\!\approx\!2.13^{\circ}$). The blue circles, squares, diamonds, and plus signs represent the microscopic configurations with $D_6$ symmetry, $D_3$ symmetry, $D_3$ symmetry with staggered sublattice potential $V_s\!=\!0.04\,$eV, and $D_3$ symmetry with vertical electric field $V_e\!=\!0.04\,$eV respectively. (b) A schematic illustration of the different microscopic configurations.   }
\label{fig3}
\end{figure}

The microscopic symmetry group of twisted bilayer graphene depends on the stacking pattern and the choice of the rotation center. For example, if before the rotation the top layer is exactly stacked on top of the bottom layer, and one takes the  center of the hexagon as the rotation center, then the resulted moir\'e superlattice has the highest symmetry $D_6$ as considered by Song \textit{et al.} \cite{song-tbg-18}. If the rotation center is chosen at one of the carbon atoms, then the resulted moir\'e structure has a $D_3$ symmetry, which is the case considered in most of the previous literatures. If the rotation center is chosen at an arbitrary point then the only symmetry the system has is $C_{2y}$ where the $``y"$ axis is along one of the morie lattice vectors.  On the other hand, the two flat bands from each valley at small twist angles have been shown to be equivalent to two zeroth Landau levels with opposite Chern numbers. The pseudo magnetic fields would lead to a new magnetic length scale $l_B$ given by Eq.~(\ref{eq:wc-lb}). Clearly $l_B$ is much greater than the microscopic lattice constant for small twist angles, thus one expects that the topological properties of the zeroth pseudo LLs should be robust regardless the perturbations on the microscopic scale.

Using the microscopic tight-binding model introduced in Eq.~(\ref{eq:hopping}), we would like to explicitly demonstrate that the topological character of the Wilson loops of the four low-energy bands remains robust regardless the microscopic details. In particular we have considered  different microscopic symmetries in the tight-binding model as schematically shown in Fig.~\ref{fig3}(b): (\i) the $D_6$ symmetry where the two layers are first stacked exactly on top of each other then rotated about the center of the hexagon; (\ii) the $D_3$ symmetry where the rotation center is at one of the carbon atoms instead of at the hexagon center; (\iii) based on the $D_3$ configuration we apply a staggered potential $V_s\!=\!0.04\,$eV on A and B sublattices in both layers; (\iv) still based on the $D_3$ configuration we apply a vertical electric field with energy $V_e\!=\!0.04\,$eV,  which breaks the $D_3$ symmetry to $C_3$ symmetry.  The Wilson loops of the four low-energy bands at $m=15$ with these four different microscopic configurations are presented in Fig.~\ref{fig3}(a), and are represented by blue circles, squares, diamonds and plus signs respectively. Clearly the Wilson loops with different microscopic symmetries almost exactly overlap with each other, indicating that the topological character of the four low-energy bands is robust against perturbations on the microscopic scale.

\subsection{The implications on the correlated insulating phases}
The pseudo-LL representation of TBG has significant implications on the correlated insulating phases observed at 1/4, 1/2, and 7/8 fillings \cite{cao-nature18-mott, choi-tbg-stm, kerelsky-tbg-stm, sharpe-tbg-19, comment_filling_tbg}. We have shown that the flat bands around the magic angles in TBG are equivalent to four zeroth Landau levels of Dirac fermions contributed by the two valleys $K$ and $K'$. The direct couplings between the two zeroth LLs (in each valley) vanish due to an emergent chiral symmetry in the zeroth LL subspace.  These four zeroth LLs carry different Chern numbers ($C$) and sublattice polarizations ($s$). In particular, from the monolayer $K$ valley the two zeroth LLs are $\{C\!=\!+1, s\!=\!A\}$, and $\{C\!=\!-1,s=\!B\}$; while for the $K'$ valley, the two zeoroth LLs have $\{C\!=\!+1, s\!=\!B\}$, $\{C\!=\!-1,s=\!A\!\}$.  Since the kinetic energy is completely quenched in the LL, any Coulomb interactions are expected to split the eight-fold (including physical spin) degenerate zeroth pseudo LLs, which would possibly lead to insulating phases at any filling factor that could completely fill up an integer number of LLs, namely at 1/8,2/8,3/8,4/8,5/8,6/8 and 7/8 fillings of the low-energy bands. Since each pseudo LL carries non-vanishing Chern number $\pm 1$, it is then quite natural that many of these phases could have nonzero total Chern number as suggested by the recent experiment on the possible quantum anomalous Hall effect in TBG \cite{sharpe-tbg-19}.

In realistic situations, the Coulomb interaction is  dependent on the layer and sublattice degrees of freedom in graphene. Transforming to the zeroth pseudo LL basis (Eq.~(\ref{eq:eigenfunction})), it means that the interaction would become dependent on the Chern number ($C$) and sublattice ($s$) polarizations of the zeroth pseudo LLs. The $\{C,s\}$ dependence of the Coulomb interaction would break the degeneracy of the insulating states at the integer fillings. The unambiguous determination of the correlated insulating ground states at the different fillings requires a microscopic and self-consistent calculation, which is beyond the scope of the present paper and we will leave it for future study.

\section{Topology of the high-energy bands} 
\label{sec:high-energy}

\subsection{Berry phases of the high-energy bands}
\begin{figure}
\includegraphics[width=2.7in]{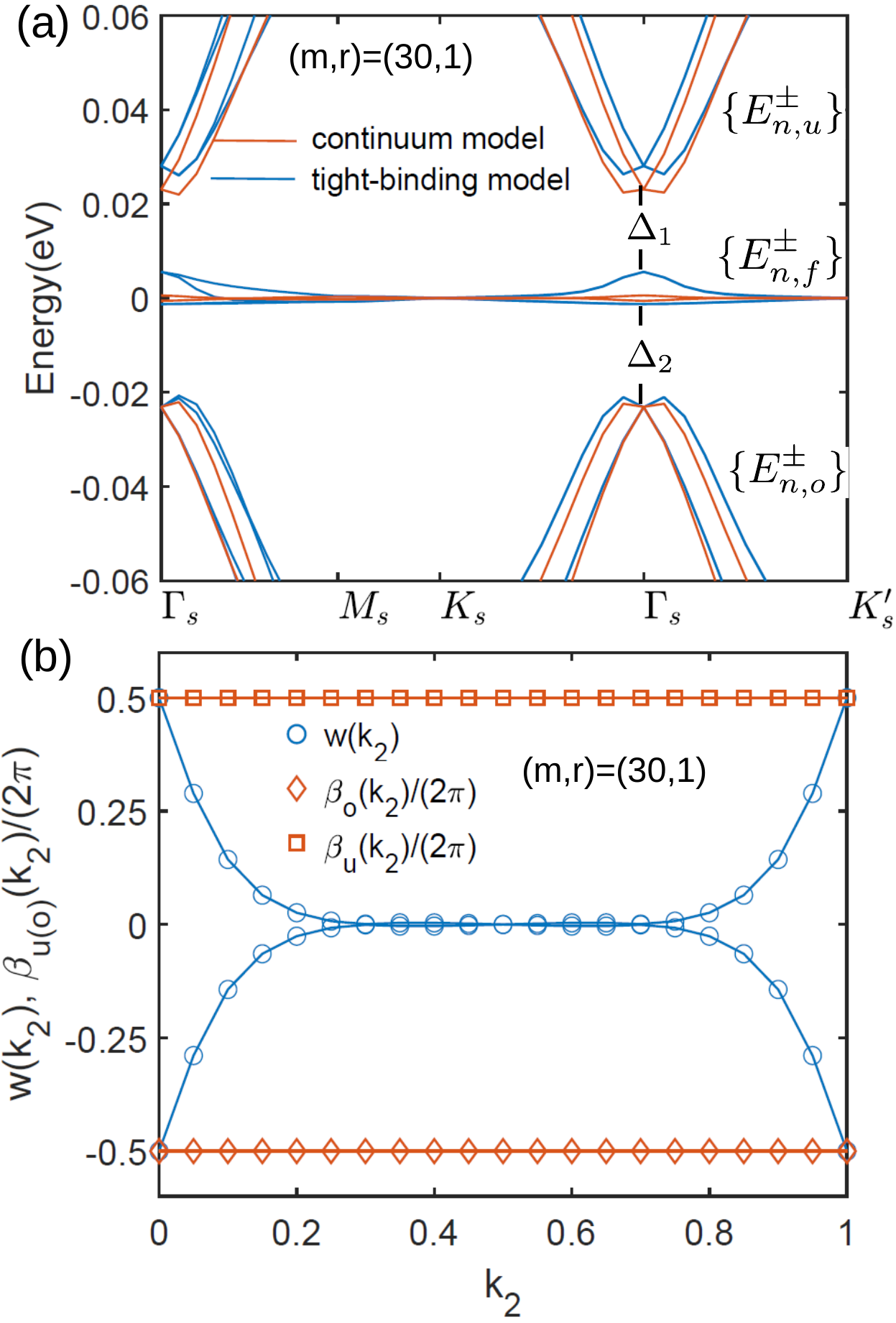}
\caption{(a) Bulk band structure of TBG at $(m,r)=(30,1)$ calculated from the microscopic tight-binding model (blue) and the continuum model (red) including effects of atomic corrugations. (b) The total Berry phases of all the bands below (above) the four flat bands at $(m,r)=(30,1)$ are denoted by $\beta_o(k_2)$ ($\beta_u(k_2)$), and the Wilson loops of the four flat bands at $(m,r)=(30,1)$ denoted by $w(k_2)$. }
\label{fig4}
\end{figure}

We continue to discuss the topological properties of the high-energy bands. We first introduce the bulk band structures before demonstrating the topology of the high-energy bands.
The bulk band structures  at $m\!=\!30$ ($\theta\!\approx\!1.08^{\circ}$) including effects of atomic corrugations  are shown in Fig.~(\ref{fig4})(a). The blue and red lines indicate the band structures calculated using the microscopic tight-binding model (Eq.~(\ref{eq:hopping})) and the continuum model (Eq.~(\ref{eq:h+})) respectively. The calculated gaps at $\Gamma_s$ above and below the four flat bands (denoted as $\Delta_1$ and $\Delta_2$ in Fig.~\ref{fig2}(a)) are around 25\,meV， which are in qualitative agreement with the experimental data \cite{cao-nature18-mott} and the theoretical calculations with fully relaxed structures \cite{koshino-tbg-prb17,tbg-relaxstructure-arxiv19}. 

Such large band gaps actually originate from the atomic corrugations: the  \textit{intersublattice} interlayer tunneling $u_0'$ generates opposite effective magnetic fields which tend to create the topological gaps; while the \textit{intrasublattice} interlayer tunneling $u_0$ tend to couple the zeroth pseudo LL in one diagonal block to the higher pseudo LLs of the other diagonal block, which reduces the topological gaps. Therefore  the gaps between the low-energy bands and high-energy bands would increase due to the atomic corrugations, because the ratio $u_0/u_0'$ decreases as a result of the atomic corrugations \cite{koshino-prx18}. This also implies that the topological properties and electronic structures of TBG can be significantly engineered using atomic corrugations, which we will discuss in details in Sec.~\ref{sec:corrugation}.

For clarity's sake we divide all the energy bands in TBG into three groups: all the bands below and above the four low-energy bands are denoted by $\{E_{n,o}^{\pm}\}$ and $\{E_{n,u}^{\pm}\}$ respectively, and the four low energy bands are denoted by $\{E_{n,f}^{\pm}\}$, where the superscripts ``+" and ``-" are indices for the two monolayer valleys $K$ and $K'$.
In addition to the four low energy bands, we find that the high energy bands $\{E_{n,o}^{\pm}\}$ and $\{E_{n,u}^{\pm}\}$ are also topologically nontrivial. To be  specific, if the Berry phase integrated along the $k_1$ direction for  the $n$th band in the group of $\{E_{n,o(u)}^{\pm}\}$  is denoted by $\beta_{n,o(u)}^{\pm}(k_2)$, then
\begin{align}
&\beta_{o}^{\pm}(k_2)\equiv\sum_{n}\beta_{n,o}^{\pm}(k_2)=\mp\pi\;,\nn
&\beta_u^{\pm}(k_2)\equiv\sum_{n}\beta_{n,u}^{\pm}(k_2)=\pm\pi\;.
\label{eq:berry-phase}
\end{align}
The quantization of the total Berry phases shown in Eq.~(\ref{eq:berry-phase}) is guaranteed by the $C_{2z}\mathcal{T}$ symmetry of the continuum model. 
Eq.~(\ref{eq:berry-phase}) has been numerically verified by implementing the continuum model of TBG in the plane-wave basis \cite{supp_info}, and are plotted in Fig.~(\ref{fig4})(b) for $m\!=\!30$ ($\theta=1.08^{\circ}$), where the red diamonds and squares represent $\beta^{+}_o(k_2)$ and $\beta^{+}_u(k_2)$ respectively. For completeness,  in Fig.~\ref{fig4}(b) we also plot the Wilson loops of the two flat bands (for one valley) as marked by the blue  circles. Thus Fig.~\ref{fig4}(b) presents the complete band topology of TBG. 

\subsection{Truncation dependence of the edge states}

\begin{figure}
\includegraphics[width=3.5in]{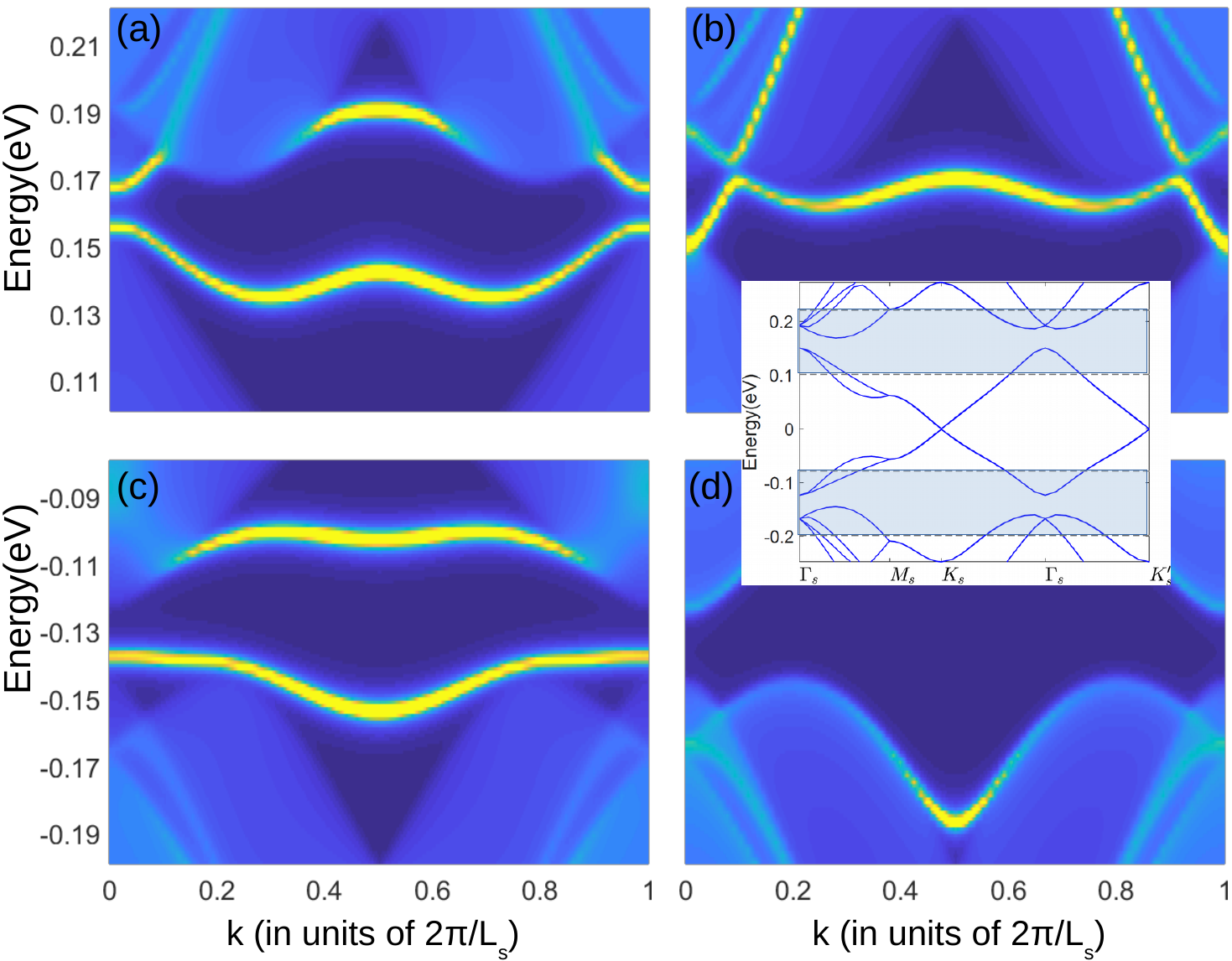}
\caption{Edge states in twist bilayer graphene at $m\!=\!15$: (a)-(b) the edge states below the four low-energy bands, and  (c)-(d) above the four low-energy bands. (a) and (c), the system truncated through the AA region; (b) and (d)), the system truncated through the BA and AB regions. The inset shows the bulk band structure for $(m,r)=(15,1)$, where energy windows for the edge states are marked in light blue shadow.}
\label{fig5}
\end{figure}

The nontrivial Berry phase of the high-energy bands implies that for each valley there would be an edge state extending through the 1D edge Brillouin zone. Moreover, because the $+$ and $-$ valleys are mapped to each other by the particle-hole-like operation as shown in Eq.~(\ref{eq:ph-symm}),   
the two edge states contributed by the two valleys may occur in the gaps below and above the four low-energy bands respectively.
In Fig.~\ref{fig4}(a) and (c) we plot the spectral functions at the edge of TBG at $m\!=\!15$ ($\theta\!\approx\!2.13^{\circ}$) in the band gaps below and above the four low-energy bands.  
Clearly there are two nearly flat edge states in the  gaps above and below  the four low-energy bands contributed by the two valleys. 
The edge states shown in Fig.~\ref{fig4}(a) and (c) are calculated when the system is truncated through the $AA$ region. If instead the system is truncated through the $BA$ and $AB$ region, then the two nearly flat edge states in the charge gaps  disappear, as shown in Fig.~\ref{fig4}(b) and (d). The truncation dependence of the edge states is reminiscent of  the property of the 1D Su-Schrieffer-Heeger (SSH) chain  with quantized Berry phase $\pm\pi$ \cite{ssh-1,vanderbilt-prb93}.
The difference is that the truncation dependence of the edge states of the TBG system occurs on the moir\'e length scale (instead of of the microscopic lattice scale), and the edge states are present when the system is truncated through the $AA$ region regardless of the orientation of the edge. For comparison we also plot the corresponding bulk bandstructure  in the inset of Fig.~\ref{fig4} where the shaded regions mark the energy windows for the edge-state calculations.
It worth to note that the edge states at $\theta\!\approx\!1.08^{\circ}$  shown in Fig.~\ref{fig2} would smoothly evolve to those shown in Fig.~\ref{fig4}(a) and (c) ($\theta\!\approx\!2.13^{\circ}$) if one of the two helical edge states gets merged into the bulk bands, leaving the other one in the bulk gap. This indicates the consistency between the high-energy and low-energy band topology.

\section{Corrugation-enhanced topological gaps and topological transitions}
\label{sec:corrugation}

In this section we study  in detail how atomic corrugations affect the electronic structures and topological properties of TBG.  We have  numerically checked that once the atomic corrugations are taken into account, the four low-energy bands are separated from the other bands by non-vanishing direct gaps  from $m\!=\!5$  all the way to $m\!=\!33$ \cite{supp_info}. It implies that the system remains topologically nontrivial for all these twist angles. The edge states predicted above thus may be a strong evidence of the nontrivial band topology in TBG.

\begin{figure}
\includegraphics[width=3.5in]{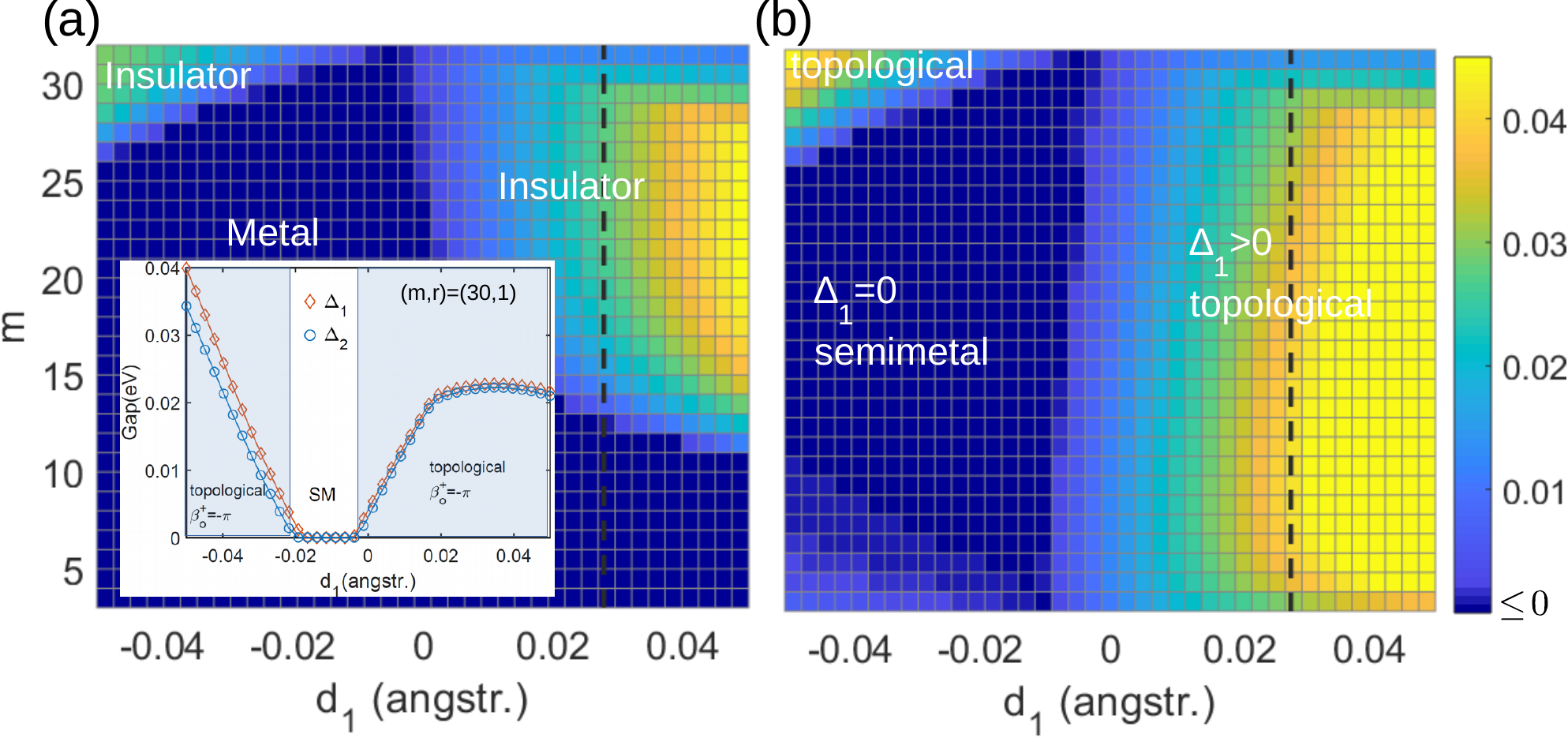}
\caption{(a)Color map of the indirect gap above the four flat bands in TBG, in units of eV. The horizontal axis is the corrugation strength parameterized by $d_1$ (see Eq.~(\ref{eq:dz-1})), and the vertical axis is the integer $m$ characterizing the rotation angle. (b) Color map of the direct gap above the four flat bands at $\Gamma_s$. The dashed black lines in (a) and (b) mark the actual atomic corrugation strength.}
\label{fig6}
\end{figure}

We further explore how the topological gaps are dependent on the corrugation strength parameterized by $d_1$ (see Eq.~(\ref{eq:dz-1})) and the twist angle $\theta(m)$.
In Fig.~\ref{fig6}(a) we plot the  the indirect gap between $\{E^{\pm}_{n,f}\}$ and $\{E^{\pm}_{n,u}\}$ in the parameter space spanned by $d_1$ and the integer $m$.
The horizontal axis is $d_1$ ranging from -0.05\,\angstrom\ to 0.05\,\angstrom, and the vertical axis is $m$, where $m\!=\!5$ corresponds to $\theta\!\approx\!6.01^{\circ}$ and $m\!=\!33$ corresponds to $\theta\!\approx\!1.02^{\circ}$.
When $m\!\leq\!11$ ($\theta\!\geq\!2.88^{\circ}$) the system is always metallic for $-0.05\,\angstrom\!\leq\!d_1\!\leq\!0.05\,$\angstrom.
When $m\!\geq\!12$ a global gap opens up as $d_1$ increases from $-0.05$\,\angstrom, indicating a transition from a metallic to a topologically nontrivial insulating phase. When $m\!\approx\!30$ ($\theta\!\approx\!1.08^{\circ}$), the system is mostly insulating and only becomes (semi)metallic in a small window of $d_1$.

In Fig.~\ref{fig6}(b) we plot the direct gap above the four low-energy bands at $\Gamma_s$ as denoted by $\Delta_1$ in Fig.~\ref{fig4}(a).  We see that when $m\!\lessapprox\!26$
$\Delta_1$ is almost exactly zero for $-0.05\,\angstrom\leq\!d_1\!\lessapprox\!0$;  when $m\!\gtrapprox\!26$ $\Delta_1$ gradually diminishes as $d_1$ decreases from 0.05\,\angstrom,  vanishes  at some critical value $d_{1c}$, then reopens at another critical value  $d_{1c}^*$. It is interesting to note that the topological character of both the four low-energy bands and the high-energy bands are \textit{unchanged} after such band-touching events at $\Gamma_s$. This is because there is (approximate) particle-hole symmetry in the low energy spectrum at $\Gamma_s$, such that the band touchings almost occur simultaneously for $\Delta_1$ and $\Delta_2$, but the band topology is not expected to be changed   after an even number of band-touching events.  This is clearly shown in the inset of Fig.~\ref{fig6}(a), where we plot $\Delta_1$ (red diamonds) and $\Delta_2$ (blue circles) as a function of $d_1$ for $m\!=\!30$.

\section{Summary} 
\label{sec:summary}
To summarize, in this paper we have proved that the two flat bands (per valley) near the magic angles in TBG originate from the two zeroth Landau levels of Dirac fermions threaded by opposite effective magnetic fields generated by the moir\'e pattern.  
The direct coupling between the two zeroth LLs is forbidden by an emergent chiral symmetry in the low-energy subspace. 
As a consequence, the two flat bands  possess opposite Chern numbers $\pm1$ and exhibit the odd winding pattern in the Wilson loops.  
This gives rise to two pairs of helical edge states in the bulk gaps between the low-energy and the high-energy bands at the first magic angle. The pseudo Landau-level representation of the flat bands in TBG have significant implications on the correlated insulating phase observed in experiments. We have argued that Coulomb interactions may split the eight-fold degenerate (including physical spin) zeroth pseudo LLs, and possibly lead to insulating states with polarized pseudo LLs with non-vanishing Chern numbers when an integer number of the pseudo LLs are filled up. 
  
We have further shown that the high-energy bands of TBG are topologically nontrivial which are characterized by constant and quantized Berry phases $\pm\pi$, and protected by $C_{2z}\mathcal{T}$ symmetry. The quantized Berry phases give rise to a pair of nearly flat edge states in the energy gaps below and above the four low-energy bands.  These edge states are robust regardless of the orientation of the edge but are dependent on the truncations on the moir\'e length scale. We also find that the topologically nontrivial gaps between the flat bands and the high-energy bands are significantly enhanced due to atomic corrugations.  Our work is a step forward in understanding the electronic properties of TBG, and have significant implications on the correlated insulating phase and superconductivity observed in TBG.



\acknowledgements
J-P.L. acknowledges the support from the Center for Scientific Computing at the California NanoSystems Institute and Materials Research Laboratory, an NSF MRSEC (DMR1720256). J-W.L. acknowledges financial support from the Hong Kong Research Grants Council (Project No. ECS26302118). X.D. acknowledges  financial support from the Hong Kong Research Grants Council (Project No. GRF16300918). We thank Zhida Song, Haizhou Lu and Shiang Fang for helpful discussions.

\bibliography{tbg}

\begin{thebibliography}{53}
\expandafter\ifx\csname natexlab\endcsname\relax\def\natexlab#1{#1}\fi
\expandafter\ifx\csname bibnamefont\endcsname\relax
  \def\bibnamefont#1{#1}\fi
\expandafter\ifx\csname bibfnamefont\endcsname\relax
  \def\bibfnamefont#1{#1}\fi
\expandafter\ifx\csname citenamefont\endcsname\relax
  \def\citenamefont#1{#1}\fi
\expandafter\ifx\csname url\endcsname\relax
  \def\url#1{\texttt{#1}}\fi
\expandafter\ifx\csname urlprefix\endcsname\relax\def\urlprefix{URL }\fi
\providecommand{\bibinfo}[2]{#2}
\providecommand{\eprint}[2][]{\url{#2}}

\bibitem[{\citenamefont{Lopes~dos Santos et~al.}(2007)\citenamefont{Lopes~dos
  Santos, Peres, and Castro~Neto}}]{santos-tbg-prl07}
\bibinfo{author}{\bibfnamefont{J.~M.~B.} \bibnamefont{Lopes~dos Santos}},
  \bibinfo{author}{\bibfnamefont{N.~M.~R.} \bibnamefont{Peres}},
  \bibnamefont{and} \bibinfo{author}{\bibfnamefont{A.~H.}
  \bibnamefont{Castro~Neto}}, \bibinfo{journal}{Phys. Rev. Lett.}
  \textbf{\bibinfo{volume}{99}}, \bibinfo{pages}{256802}
  (\bibinfo{year}{2007}).

\bibitem[{\citenamefont{Bistritzer and MacDonald}(2011)}]{macdonald-pnas11}
\bibinfo{author}{\bibfnamefont{R.}~\bibnamefont{Bistritzer}} \bibnamefont{and}
  \bibinfo{author}{\bibfnamefont{A.~H.} \bibnamefont{MacDonald}},
  \bibinfo{journal}{Proceedings of the National Academy of Sciences}
  \textbf{\bibinfo{volume}{108}}, \bibinfo{pages}{12233}
  (\bibinfo{year}{2011}).

\bibitem[{\citenamefont{Li et~al.}(2010)\citenamefont{Li, Luican, Dos~Santos,
  Neto, Reina, Kong, and Andrei}}]{li-vhs-11}
\bibinfo{author}{\bibfnamefont{G.}~\bibnamefont{Li}},
  \bibinfo{author}{\bibfnamefont{A.}~\bibnamefont{Luican}},
  \bibinfo{author}{\bibfnamefont{J.~L.} \bibnamefont{Dos~Santos}},
  \bibinfo{author}{\bibfnamefont{A.~C.} \bibnamefont{Neto}},
  \bibinfo{author}{\bibfnamefont{A.}~\bibnamefont{Reina}},
  \bibinfo{author}{\bibfnamefont{J.}~\bibnamefont{Kong}}, \bibnamefont{and}
  \bibinfo{author}{\bibfnamefont{E.}~\bibnamefont{Andrei}},
  \bibinfo{journal}{Nature Physics} \textbf{\bibinfo{volume}{6}},
  \bibinfo{pages}{109} (\bibinfo{year}{2010}).

\bibitem[{\citenamefont{Lee et~al.}(2011)\citenamefont{Lee, Riedl, Beringer,
  Castro~Neto, von Klitzing, Starke, and Smet}}]{smet-qhe-tbg-prl11}
\bibinfo{author}{\bibfnamefont{D.~S.} \bibnamefont{Lee}},
  \bibinfo{author}{\bibfnamefont{C.}~\bibnamefont{Riedl}},
  \bibinfo{author}{\bibfnamefont{T.}~\bibnamefont{Beringer}},
  \bibinfo{author}{\bibfnamefont{A.~H.} \bibnamefont{Castro~Neto}},
  \bibinfo{author}{\bibfnamefont{K.}~\bibnamefont{von Klitzing}},
  \bibinfo{author}{\bibfnamefont{U.}~\bibnamefont{Starke}}, \bibnamefont{and}
  \bibinfo{author}{\bibfnamefont{J.~H.} \bibnamefont{Smet}},
  \bibinfo{journal}{Phys. Rev. Lett.} \textbf{\bibinfo{volume}{107}},
  \bibinfo{pages}{216602} (\bibinfo{year}{2011}).

\bibitem[{\citenamefont{Sanchez-Yamagishi
  et~al.}(2012)\citenamefont{Sanchez-Yamagishi, Taychatanapat, Watanabe,
  Taniguchi, Yacoby, and Jarillo-Herrero}}]{qhe-tbg-prl12}
\bibinfo{author}{\bibfnamefont{J.~D.} \bibnamefont{Sanchez-Yamagishi}},
  \bibinfo{author}{\bibfnamefont{T.}~\bibnamefont{Taychatanapat}},
  \bibinfo{author}{\bibfnamefont{K.}~\bibnamefont{Watanabe}},
  \bibinfo{author}{\bibfnamefont{T.}~\bibnamefont{Taniguchi}},
  \bibinfo{author}{\bibfnamefont{A.}~\bibnamefont{Yacoby}}, \bibnamefont{and}
  \bibinfo{author}{\bibfnamefont{P.}~\bibnamefont{Jarillo-Herrero}},
  \bibinfo{journal}{Phys. Rev. Lett.} \textbf{\bibinfo{volume}{108}},
  \bibinfo{pages}{076601} (\bibinfo{year}{2012}).

\bibitem[{\citenamefont{Yan et~al.}(2012)\citenamefont{Yan, Liu, Dou, Meng,
  Feng, Chu, Zhang, Liu, Nie, and He}}]{he-prl11}
\bibinfo{author}{\bibfnamefont{W.}~\bibnamefont{Yan}},
  \bibinfo{author}{\bibfnamefont{M.}~\bibnamefont{Liu}},
  \bibinfo{author}{\bibfnamefont{R.-F.} \bibnamefont{Dou}},
  \bibinfo{author}{\bibfnamefont{L.}~\bibnamefont{Meng}},
  \bibinfo{author}{\bibfnamefont{L.}~\bibnamefont{Feng}},
  \bibinfo{author}{\bibfnamefont{Z.-D.} \bibnamefont{Chu}},
  \bibinfo{author}{\bibfnamefont{Y.}~\bibnamefont{Zhang}},
  \bibinfo{author}{\bibfnamefont{Z.}~\bibnamefont{Liu}},
  \bibinfo{author}{\bibfnamefont{J.-C.} \bibnamefont{Nie}}, \bibnamefont{and}
  \bibinfo{author}{\bibfnamefont{L.}~\bibnamefont{He}}, \bibinfo{journal}{Phys.
  Rev. Lett.} \textbf{\bibinfo{volume}{109}}, \bibinfo{pages}{126801}
  (\bibinfo{year}{2012}).

\bibitem[{\citenamefont{San-Jose et~al.}(2012)\citenamefont{San-Jose,
  Gonz\'alez, and Guinea}}]{tbg-nonabelian-prl12}
\bibinfo{author}{\bibfnamefont{P.}~\bibnamefont{San-Jose}},
  \bibinfo{author}{\bibfnamefont{J.}~\bibnamefont{Gonz\'alez}},
  \bibnamefont{and} \bibinfo{author}{\bibfnamefont{F.}~\bibnamefont{Guinea}},
  \bibinfo{journal}{Phys. Rev. Lett.} \textbf{\bibinfo{volume}{108}},
  \bibinfo{pages}{216802} (\bibinfo{year}{2012}).

\bibitem[{\citenamefont{Cao et~al.}(2018{\natexlab{a}})\citenamefont{Cao,
  Fatemi, Demir, Fang, Tomarken, Luo, Sanchez-Yamagishi, Watanabe, Taniguchi,
  Kaxiras et~al.}}]{cao-nature18-mott}
\bibinfo{author}{\bibfnamefont{Y.}~\bibnamefont{Cao}},
  \bibinfo{author}{\bibfnamefont{V.}~\bibnamefont{Fatemi}},
  \bibinfo{author}{\bibfnamefont{A.}~\bibnamefont{Demir}},
  \bibinfo{author}{\bibfnamefont{S.}~\bibnamefont{Fang}},
  \bibinfo{author}{\bibfnamefont{S.~L.} \bibnamefont{Tomarken}},
  \bibinfo{author}{\bibfnamefont{J.~Y.} \bibnamefont{Luo}},
  \bibinfo{author}{\bibfnamefont{J.~D.} \bibnamefont{Sanchez-Yamagishi}},
  \bibinfo{author}{\bibfnamefont{K.}~\bibnamefont{Watanabe}},
  \bibinfo{author}{\bibfnamefont{T.}~\bibnamefont{Taniguchi}},
  \bibinfo{author}{\bibfnamefont{E.}~\bibnamefont{Kaxiras}},
  \bibnamefont{et~al.}, \bibinfo{journal}{Nature}
  \textbf{\bibinfo{volume}{556}}, \bibinfo{pages}{80}
  (\bibinfo{year}{2018}{\natexlab{a}}).

\bibitem[{\citenamefont{Sharpe et~al.}(2019)\citenamefont{Sharpe, Fox, Barnard,
  Finney, Watanabe, Taniguchi, Kastner, and Goldhaber-Gordon}}]{sharpe-tbg-19}
\bibinfo{author}{\bibfnamefont{A.~L.} \bibnamefont{Sharpe}},
  \bibinfo{author}{\bibfnamefont{E.~J.} \bibnamefont{Fox}},
  \bibinfo{author}{\bibfnamefont{A.~W.} \bibnamefont{Barnard}},
  \bibinfo{author}{\bibfnamefont{J.}~\bibnamefont{Finney}},
  \bibinfo{author}{\bibfnamefont{K.}~\bibnamefont{Watanabe}},
  \bibinfo{author}{\bibfnamefont{T.}~\bibnamefont{Taniguchi}},
  \bibinfo{author}{\bibfnamefont{M.}~\bibnamefont{Kastner}}, \bibnamefont{and}
  \bibinfo{author}{\bibfnamefont{D.}~\bibnamefont{Goldhaber-Gordon}},
  \bibinfo{journal}{arXiv preprint arXiv:1901.03520}  (\bibinfo{year}{2019}).

\bibitem[{\citenamefont{Choi et~al.}(2019)\citenamefont{Choi, Kemmer, Peng,
  Thomson, Arora, Polski, Zhang, Ren, Alicea, Refael et~al.}}]{choi-tbg-stm}
\bibinfo{author}{\bibfnamefont{Y.}~\bibnamefont{Choi}},
  \bibinfo{author}{\bibfnamefont{J.}~\bibnamefont{Kemmer}},
  \bibinfo{author}{\bibfnamefont{Y.}~\bibnamefont{Peng}},
  \bibinfo{author}{\bibfnamefont{A.}~\bibnamefont{Thomson}},
  \bibinfo{author}{\bibfnamefont{H.}~\bibnamefont{Arora}},
  \bibinfo{author}{\bibfnamefont{R.}~\bibnamefont{Polski}},
  \bibinfo{author}{\bibfnamefont{Y.}~\bibnamefont{Zhang}},
  \bibinfo{author}{\bibfnamefont{H.}~\bibnamefont{Ren}},
  \bibinfo{author}{\bibfnamefont{J.}~\bibnamefont{Alicea}},
  \bibinfo{author}{\bibfnamefont{G.}~\bibnamefont{Refael}},
  \bibnamefont{et~al.}, \bibinfo{journal}{arXiv preprint arXiv:1901.02997}
  (\bibinfo{year}{2019}).

\bibitem[{\citenamefont{Kerelsky et~al.}(2018)\citenamefont{Kerelsky, McGilly,
  Kennes, Xian, Yankowitz, Chen, Watanabe, Taniguchi, Hone, Dean
  et~al.}}]{kerelsky-tbg-stm}
\bibinfo{author}{\bibfnamefont{A.}~\bibnamefont{Kerelsky}},
  \bibinfo{author}{\bibfnamefont{L.}~\bibnamefont{McGilly}},
  \bibinfo{author}{\bibfnamefont{D.~M.} \bibnamefont{Kennes}},
  \bibinfo{author}{\bibfnamefont{L.}~\bibnamefont{Xian}},
  \bibinfo{author}{\bibfnamefont{M.}~\bibnamefont{Yankowitz}},
  \bibinfo{author}{\bibfnamefont{S.}~\bibnamefont{Chen}},
  \bibinfo{author}{\bibfnamefont{K.}~\bibnamefont{Watanabe}},
  \bibinfo{author}{\bibfnamefont{T.}~\bibnamefont{Taniguchi}},
  \bibinfo{author}{\bibfnamefont{J.}~\bibnamefont{Hone}},
  \bibinfo{author}{\bibfnamefont{C.}~\bibnamefont{Dean}}, \bibnamefont{et~al.},
  \bibinfo{journal}{arXiv preprint arXiv:1812.08776}  (\bibinfo{year}{2018}).

\bibitem[{\citenamefont{Cao et~al.}(2018{\natexlab{b}})\citenamefont{Cao,
  Fatemi, Fang, Watanabe, Taniguchi, Kaxiras, and
  Jarillo-Herrero}}]{cao-nature18-supercond}
\bibinfo{author}{\bibfnamefont{Y.}~\bibnamefont{Cao}},
  \bibinfo{author}{\bibfnamefont{V.}~\bibnamefont{Fatemi}},
  \bibinfo{author}{\bibfnamefont{S.}~\bibnamefont{Fang}},
  \bibinfo{author}{\bibfnamefont{K.}~\bibnamefont{Watanabe}},
  \bibinfo{author}{\bibfnamefont{T.}~\bibnamefont{Taniguchi}},
  \bibinfo{author}{\bibfnamefont{E.}~\bibnamefont{Kaxiras}}, \bibnamefont{and}
  \bibinfo{author}{\bibfnamefont{P.}~\bibnamefont{Jarillo-Herrero}},
  \bibinfo{journal}{Nature} \textbf{\bibinfo{volume}{556}}, \bibinfo{pages}{43}
  (\bibinfo{year}{2018}{\natexlab{b}}).

\bibitem[{\citenamefont{Po et~al.}(2018{\natexlab{a}})\citenamefont{Po, Zou,
  Vishwanath, and Senthil}}]{po-prx18}
\bibinfo{author}{\bibfnamefont{H.~C.} \bibnamefont{Po}},
  \bibinfo{author}{\bibfnamefont{L.}~\bibnamefont{Zou}},
  \bibinfo{author}{\bibfnamefont{A.}~\bibnamefont{Vishwanath}},
  \bibnamefont{and} \bibinfo{author}{\bibfnamefont{T.}~\bibnamefont{Senthil}},
  \bibinfo{journal}{Phys. Rev. X} \textbf{\bibinfo{volume}{8}},
  \bibinfo{pages}{031089} (\bibinfo{year}{2018}{\natexlab{a}}).

\bibitem[{\citenamefont{Yuan and Fu}(2018)}]{yuan-prb18}
\bibinfo{author}{\bibfnamefont{N.~F.~Q.} \bibnamefont{Yuan}} \bibnamefont{and}
  \bibinfo{author}{\bibfnamefont{L.}~\bibnamefont{Fu}}, \bibinfo{journal}{Phys.
  Rev. B} \textbf{\bibinfo{volume}{98}}, \bibinfo{pages}{045103}
  (\bibinfo{year}{2018}).

\bibitem[{\citenamefont{Koshino et~al.}(2018)\citenamefont{Koshino, Yuan,
  Koretsune, Ochi, Kuroki, and Fu}}]{koshino-prx18}
\bibinfo{author}{\bibfnamefont{M.}~\bibnamefont{Koshino}},
  \bibinfo{author}{\bibfnamefont{N.~F.~Q.} \bibnamefont{Yuan}},
  \bibinfo{author}{\bibfnamefont{T.}~\bibnamefont{Koretsune}},
  \bibinfo{author}{\bibfnamefont{M.}~\bibnamefont{Ochi}},
  \bibinfo{author}{\bibfnamefont{K.}~\bibnamefont{Kuroki}}, \bibnamefont{and}
  \bibinfo{author}{\bibfnamefont{L.}~\bibnamefont{Fu}}, \bibinfo{journal}{Phys.
  Rev. X} \textbf{\bibinfo{volume}{8}}, \bibinfo{pages}{031087}
  (\bibinfo{year}{2018}).

\bibitem[{\citenamefont{Kang and Vafek}(2018{\natexlab{a}})}]{kang-prx18}
\bibinfo{author}{\bibfnamefont{J.}~\bibnamefont{Kang}} \bibnamefont{and}
  \bibinfo{author}{\bibfnamefont{O.}~\bibnamefont{Vafek}},
  \bibinfo{journal}{Phys. Rev. X} \textbf{\bibinfo{volume}{8}},
  \bibinfo{pages}{031088} (\bibinfo{year}{2018}{\natexlab{a}}).

\bibitem[{\citenamefont{Song et~al.}(2018)\citenamefont{Song, Wang, Shi, Li,
  Fang, and Bernevig}}]{song-tbg-18}
\bibinfo{author}{\bibfnamefont{Z.}~\bibnamefont{Song}},
  \bibinfo{author}{\bibfnamefont{Z.}~\bibnamefont{Wang}},
  \bibinfo{author}{\bibfnamefont{W.}~\bibnamefont{Shi}},
  \bibinfo{author}{\bibfnamefont{G.}~\bibnamefont{Li}},
  \bibinfo{author}{\bibfnamefont{C.}~\bibnamefont{Fang}}, \bibnamefont{and}
  \bibinfo{author}{\bibfnamefont{B.~A.} \bibnamefont{Bernevig}},
  \bibinfo{journal}{arXiv preprint arXiv:1807.10676}  (\bibinfo{year}{2018}).

\bibitem[{\citenamefont{Po et~al.}(2018{\natexlab{b}})\citenamefont{Po, Zou,
  Senthil, and Vishwanath}}]{po-tbg2}
\bibinfo{author}{\bibfnamefont{H.~C.} \bibnamefont{Po}},
  \bibinfo{author}{\bibfnamefont{L.}~\bibnamefont{Zou}},
  \bibinfo{author}{\bibfnamefont{T.}~\bibnamefont{Senthil}}, \bibnamefont{and}
  \bibinfo{author}{\bibfnamefont{A.}~\bibnamefont{Vishwanath}},
  \bibinfo{journal}{arXiv preprint arXiv:1808.02482}
  (\bibinfo{year}{2018}{\natexlab{b}}).

\bibitem[{\citenamefont{Hejazi et~al.}(2018)\citenamefont{Hejazi, Liu,
  Shapourian, Chen, and Balents}}]{hejazi-arxiv18}
\bibinfo{author}{\bibfnamefont{K.}~\bibnamefont{Hejazi}},
  \bibinfo{author}{\bibfnamefont{C.}~\bibnamefont{Liu}},
  \bibinfo{author}{\bibfnamefont{H.}~\bibnamefont{Shapourian}},
  \bibinfo{author}{\bibfnamefont{X.}~\bibnamefont{Chen}}, \bibnamefont{and}
  \bibinfo{author}{\bibfnamefont{L.}~\bibnamefont{Balents}},
  \bibinfo{journal}{arXiv preprint arXiv:1808.01568}  (\bibinfo{year}{2018}).

\bibitem[{\citenamefont{Tarnopolsky et~al.}(2018)\citenamefont{Tarnopolsky,
  Kruchkov, and Vishwanath}}]{origin-magic-angle-tbg18}
\bibinfo{author}{\bibfnamefont{G.}~\bibnamefont{Tarnopolsky}},
  \bibinfo{author}{\bibfnamefont{A.~J.} \bibnamefont{Kruchkov}},
  \bibnamefont{and}
  \bibinfo{author}{\bibfnamefont{A.}~\bibnamefont{Vishwanath}},
  \bibinfo{journal}{arXiv preprint arXiv:1808.05250}  (\bibinfo{year}{2018}).

\bibitem[{\citenamefont{Ramires and Lado}(2018)}]{ramires-prl18}
\bibinfo{author}{\bibfnamefont{A.}~\bibnamefont{Ramires}} \bibnamefont{and}
  \bibinfo{author}{\bibfnamefont{J.~L.} \bibnamefont{Lado}},
  \bibinfo{journal}{Phys. Rev. Lett.} \textbf{\bibinfo{volume}{121}},
  \bibinfo{pages}{146801} (\bibinfo{year}{2018}).

\bibitem[{\citenamefont{Pal et~al.}(2018)\citenamefont{Pal, Spitz, and
  Kindermann}}]{pal-kindermann-arxiv18}
\bibinfo{author}{\bibfnamefont{H.~K.} \bibnamefont{Pal}},
  \bibinfo{author}{\bibfnamefont{S.}~\bibnamefont{Spitz}}, \bibnamefont{and}
  \bibinfo{author}{\bibfnamefont{M.}~\bibnamefont{Kindermann}},
  \bibinfo{journal}{arXiv preprint arXiv:1803.07060}  (\bibinfo{year}{2018}).

\bibitem[{\citenamefont{Lian et~al.}(2018{\natexlab{a}})\citenamefont{Lian,
  Xie, and Bernevig}}]{ll-tbg-lian}
\bibinfo{author}{\bibfnamefont{B.}~\bibnamefont{Lian}},
  \bibinfo{author}{\bibfnamefont{F.}~\bibnamefont{Xie}}, \bibnamefont{and}
  \bibinfo{author}{\bibfnamefont{A.}~\bibnamefont{Bernevig}, \bibfnamefont{B}},
  \bibinfo{journal}{arXiv preprint arXiv:1811.11786}
  (\bibinfo{year}{2018}{\natexlab{a}}).

\bibitem[{\citenamefont{Nam and Koshino}(2017)}]{koshino-tbg-prb17}
\bibinfo{author}{\bibfnamefont{N.~N.~T.} \bibnamefont{Nam}} \bibnamefont{and}
  \bibinfo{author}{\bibfnamefont{M.}~\bibnamefont{Koshino}},
  \bibinfo{journal}{Phys. Rev. B} \textbf{\bibinfo{volume}{96}},
  \bibinfo{pages}{075311} (\bibinfo{year}{2017}).

\bibitem[{\citenamefont{Jain et~al.}(2016)\citenamefont{Jain,
  Juri{\v{c}}i{\'c}, and Barkema}}]{jain-tbg-structure}
\bibinfo{author}{\bibfnamefont{S.~K.} \bibnamefont{Jain}},
  \bibinfo{author}{\bibfnamefont{V.}~\bibnamefont{Juri{\v{c}}i{\'c}}},
  \bibnamefont{and} \bibinfo{author}{\bibfnamefont{G.~T.}
  \bibnamefont{Barkema}}, \bibinfo{journal}{2D Materials}
  \textbf{\bibinfo{volume}{4}}, \bibinfo{pages}{015018} (\bibinfo{year}{2016}).

\bibitem[{\citenamefont{Angeli et~al.}(2018)\citenamefont{Angeli, Mandelli,
  Valli, Amaricci, Capone, Tosatti, and Fabrizio}}]{tbg-D6-arxiv18}
\bibinfo{author}{\bibfnamefont{M.}~\bibnamefont{Angeli}},
  \bibinfo{author}{\bibfnamefont{D.}~\bibnamefont{Mandelli}},
  \bibinfo{author}{\bibfnamefont{A.}~\bibnamefont{Valli}},
  \bibinfo{author}{\bibfnamefont{A.}~\bibnamefont{Amaricci}},
  \bibinfo{author}{\bibfnamefont{M.}~\bibnamefont{Capone}},
  \bibinfo{author}{\bibfnamefont{E.}~\bibnamefont{Tosatti}}, \bibnamefont{and}
  \bibinfo{author}{\bibfnamefont{M.}~\bibnamefont{Fabrizio}},
  \bibinfo{journal}{arXiv preprint arXiv:1809.11140}  (\bibinfo{year}{2018}).

\bibitem[{\citenamefont{Sboychakov et~al.}(2018)\citenamefont{Sboychakov,
  Rozhkov, Rakhmanov, and Nori}}]{sboychakov-arxiv-18}
\bibinfo{author}{\bibfnamefont{A.}~\bibnamefont{Sboychakov}},
  \bibinfo{author}{\bibfnamefont{A.}~\bibnamefont{Rozhkov}},
  \bibinfo{author}{\bibfnamefont{A.}~\bibnamefont{Rakhmanov}},
  \bibnamefont{and} \bibinfo{author}{\bibfnamefont{F.}~\bibnamefont{Nori}},
  \bibinfo{journal}{arXiv preprint arXiv:1807.08190}  (\bibinfo{year}{2018}).

\bibitem[{\citenamefont{Isobe et~al.}(2018)\citenamefont{Isobe, Yuan, and
  Fu}}]{isobe-prx18}
\bibinfo{author}{\bibfnamefont{H.}~\bibnamefont{Isobe}},
  \bibinfo{author}{\bibfnamefont{N.~F.~Q.} \bibnamefont{Yuan}},
  \bibnamefont{and} \bibinfo{author}{\bibfnamefont{L.}~\bibnamefont{Fu}},
  \bibinfo{journal}{Phys. Rev. X} \textbf{\bibinfo{volume}{8}},
  \bibinfo{pages}{041041} (\bibinfo{year}{2018}).

\bibitem[{\citenamefont{Xu et~al.}(2018)\citenamefont{Xu, Law, and
  Lee}}]{xu-lee-prb18}
\bibinfo{author}{\bibfnamefont{X.~Y.} \bibnamefont{Xu}},
  \bibinfo{author}{\bibfnamefont{K.~T.} \bibnamefont{Law}}, \bibnamefont{and}
  \bibinfo{author}{\bibfnamefont{P.~A.} \bibnamefont{Lee}},
  \bibinfo{journal}{Phys. Rev. B} \textbf{\bibinfo{volume}{98}},
  \bibinfo{pages}{121406} (\bibinfo{year}{2018}).

\bibitem[{\citenamefont{Huang et~al.}(2018)\citenamefont{Huang, Zhang, and
  Ma}}]{huang-arxiv-18}
\bibinfo{author}{\bibfnamefont{T.}~\bibnamefont{Huang}},
  \bibinfo{author}{\bibfnamefont{L.}~\bibnamefont{Zhang}}, \bibnamefont{and}
  \bibinfo{author}{\bibfnamefont{T.}~\bibnamefont{Ma}}, \bibinfo{journal}{arXiv
  preprint arXiv:1804.06096}  (\bibinfo{year}{2018}).

\bibitem[{\citenamefont{Liu et~al.}(2018)\citenamefont{Liu, Zhang, Chen, and
  Yang}}]{liu-prl18}
\bibinfo{author}{\bibfnamefont{C.-C.} \bibnamefont{Liu}},
  \bibinfo{author}{\bibfnamefont{L.-D.} \bibnamefont{Zhang}},
  \bibinfo{author}{\bibfnamefont{W.-Q.} \bibnamefont{Chen}}, \bibnamefont{and}
  \bibinfo{author}{\bibfnamefont{F.}~\bibnamefont{Yang}},
  \bibinfo{journal}{Phys. Rev. Lett.} \textbf{\bibinfo{volume}{121}},
  \bibinfo{pages}{217001} (\bibinfo{year}{2018}).

\bibitem[{\citenamefont{Rademaker and Mellado}(2018)}]{rademaker-prb18}
\bibinfo{author}{\bibfnamefont{L.}~\bibnamefont{Rademaker}} \bibnamefont{and}
  \bibinfo{author}{\bibfnamefont{P.}~\bibnamefont{Mellado}},
  \bibinfo{journal}{Phys. Rev. B} \textbf{\bibinfo{volume}{98}},
  \bibinfo{pages}{235158} (\bibinfo{year}{2018}).

\bibitem[{\citenamefont{Venderbos and Fernandes}(2018)}]{venderbos-prb18}
\bibinfo{author}{\bibfnamefont{J.~W.~F.} \bibnamefont{Venderbos}}
  \bibnamefont{and} \bibinfo{author}{\bibfnamefont{R.~M.}
  \bibnamefont{Fernandes}}, \bibinfo{journal}{Phys. Rev. B}
  \textbf{\bibinfo{volume}{98}}, \bibinfo{pages}{245103}
  (\bibinfo{year}{2018}).

\bibitem[{\citenamefont{Kang and
  Vafek}(2018{\natexlab{b}})}]{kang-tbg-correlation-arxiv18}
\bibinfo{author}{\bibfnamefont{J.}~\bibnamefont{Kang}} \bibnamefont{and}
  \bibinfo{author}{\bibfnamefont{O.}~\bibnamefont{Vafek}},
  \bibinfo{journal}{arXiv preprint arXiv:1810.08642}
  (\bibinfo{year}{2018}{\natexlab{b}}).

\bibitem[{\citenamefont{Xie and MacDonald}(2018)}]{xie-tbg-2018}
\bibinfo{author}{\bibfnamefont{M.}~\bibnamefont{Xie}} \bibnamefont{and}
  \bibinfo{author}{\bibfnamefont{A.~H.} \bibnamefont{MacDonald}},
  \bibinfo{journal}{arXiv preprint arXiv:1812.04213}  (\bibinfo{year}{2018}).

\bibitem[{\citenamefont{Bultinck et~al.}(2019)\citenamefont{Bultinck,
  Chatterjee, and Zaletel}}]{zaletel-tbg-2019}
\bibinfo{author}{\bibfnamefont{N.}~\bibnamefont{Bultinck}},
  \bibinfo{author}{\bibfnamefont{S.}~\bibnamefont{Chatterjee}},
  \bibnamefont{and} \bibinfo{author}{\bibfnamefont{M.~P.}
  \bibnamefont{Zaletel}}, \bibinfo{journal}{arXiv preprint arXiv:1901.08110}
  (\bibinfo{year}{2019}).

\bibitem[{\citenamefont{Xu and Balents}(2018)}]{xu-prl18}
\bibinfo{author}{\bibfnamefont{C.}~\bibnamefont{Xu}} \bibnamefont{and}
  \bibinfo{author}{\bibfnamefont{L.}~\bibnamefont{Balents}},
  \bibinfo{journal}{Phys. Rev. Lett.} \textbf{\bibinfo{volume}{121}},
  \bibinfo{pages}{087001} (\bibinfo{year}{2018}).

\bibitem[{\citenamefont{Wu et~al.}(2018{\natexlab{a}})\citenamefont{Wu,
  MacDonald, and Martin}}]{wu-prl18}
\bibinfo{author}{\bibfnamefont{F.}~\bibnamefont{Wu}},
  \bibinfo{author}{\bibfnamefont{A.~H.} \bibnamefont{MacDonald}},
  \bibnamefont{and} \bibinfo{author}{\bibfnamefont{I.}~\bibnamefont{Martin}},
  \bibinfo{journal}{Phys. Rev. Lett.} \textbf{\bibinfo{volume}{121}},
  \bibinfo{pages}{257001} (\bibinfo{year}{2018}{\natexlab{a}}).

\bibitem[{\citenamefont{Wu et~al.}(2018{\natexlab{b}})\citenamefont{Wu, Pawlak,
  Jian, and Xu}}]{wu-xu-arxiv-18}
\bibinfo{author}{\bibfnamefont{X.-C.} \bibnamefont{Wu}},
  \bibinfo{author}{\bibfnamefont{K.~A.} \bibnamefont{Pawlak}},
  \bibinfo{author}{\bibfnamefont{C.-M.} \bibnamefont{Jian}}, \bibnamefont{and}
  \bibinfo{author}{\bibfnamefont{C.}~\bibnamefont{Xu}}, \bibinfo{journal}{arXiv
  preprint arXiv:1805.06906}  (\bibinfo{year}{2018}{\natexlab{b}}).

\bibitem[{\citenamefont{Lian et~al.}(2018{\natexlab{b}})\citenamefont{Lian,
  Wang, and Bernevig}}]{lian-arxiv-18}
\bibinfo{author}{\bibfnamefont{B.}~\bibnamefont{Lian}},
  \bibinfo{author}{\bibfnamefont{Z.}~\bibnamefont{Wang}}, \bibnamefont{and}
  \bibinfo{author}{\bibfnamefont{B.~A.} \bibnamefont{Bernevig}},
  \bibinfo{journal}{arXiv preprint arXiv:1807.04382}
  (\bibinfo{year}{2018}{\natexlab{b}}).

\bibitem[{\citenamefont{Kozii et~al.}(2018)\citenamefont{Kozii, Isobe,
  Venderbos, and Fu}}]{nematic-tbg-arxiv18}
\bibinfo{author}{\bibfnamefont{V.}~\bibnamefont{Kozii}},
  \bibinfo{author}{\bibfnamefont{H.}~\bibnamefont{Isobe}},
  \bibinfo{author}{\bibfnamefont{J.~W.} \bibnamefont{Venderbos}},
  \bibnamefont{and} \bibinfo{author}{\bibfnamefont{L.}~\bibnamefont{Fu}},
  \bibinfo{journal}{arXiv preprint arXiv:1810.04159}  (\bibinfo{year}{2018}).

\bibitem[{\citenamefont{Wu}(2018)}]{wu-tbg-chiral-arxiv18}
\bibinfo{author}{\bibfnamefont{F.}~\bibnamefont{Wu}}, \bibinfo{journal}{arXiv
  preprint arXiv:1811.10620}  (\bibinfo{year}{2018}).

\bibitem[{\citenamefont{Ahn et~al.}(2018)\citenamefont{Ahn, Park, and
  Yang}}]{yang-tbg18}
\bibinfo{author}{\bibfnamefont{J.}~\bibnamefont{Ahn}},
  \bibinfo{author}{\bibfnamefont{S.}~\bibnamefont{Park}}, \bibnamefont{and}
  \bibinfo{author}{\bibfnamefont{B.-J.} \bibnamefont{Yang}},
  \bibinfo{journal}{arXiv preprint arXiv:1808.05375}  (\bibinfo{year}{2018}).

\bibitem[{com()}]{comment_filling_tbg}
\bibinfo{note}{In this paper we define the filling factor as $\nu/8$, where
  $\nu$ is the number of filled bands out of all the 8 low-energy bands
  (including physical spin) in TBG. This is different from the convention
  adopted in some of the previous literatures, in which the filling factor is
  defined as the filling of half of all the low-energy bands.}

\bibitem[{\citenamefont{Lopes~dos Santos et~al.}(2012)\citenamefont{Lopes~dos
  Santos, Peres, and Castro~Neto}}]{castro-neto-prb12}
\bibinfo{author}{\bibfnamefont{J.~M.~B.} \bibnamefont{Lopes~dos Santos}},
  \bibinfo{author}{\bibfnamefont{N.~M.~R.} \bibnamefont{Peres}},
  \bibnamefont{and} \bibinfo{author}{\bibfnamefont{A.~H.}
  \bibnamefont{Castro~Neto}}, \bibinfo{journal}{Phys. Rev. B}
  \textbf{\bibinfo{volume}{86}}, \bibinfo{pages}{155449}
  (\bibinfo{year}{2012}).

\bibitem[{\citenamefont{Uchida et~al.}(2014)\citenamefont{Uchida, Furuya,
  Iwata, and Oshiyama}}]{uchida-corrugation-prb14}
\bibinfo{author}{\bibfnamefont{K.}~\bibnamefont{Uchida}},
  \bibinfo{author}{\bibfnamefont{S.}~\bibnamefont{Furuya}},
  \bibinfo{author}{\bibfnamefont{J.-I.} \bibnamefont{Iwata}}, \bibnamefont{and}
  \bibinfo{author}{\bibfnamefont{A.}~\bibnamefont{Oshiyama}},
  \bibinfo{journal}{Phys. Rev. B} \textbf{\bibinfo{volume}{90}},
  \bibinfo{pages}{155451} (\bibinfo{year}{2014}).

\bibitem[{\citenamefont{Lee et~al.}(2008)\citenamefont{Lee, Lee, Ahn, Kim,
  Wilson, and John}}]{graphite-AA}
\bibinfo{author}{\bibfnamefont{J.-K.} \bibnamefont{Lee}},
  \bibinfo{author}{\bibfnamefont{S.-C.} \bibnamefont{Lee}},
  \bibinfo{author}{\bibfnamefont{J.-P.} \bibnamefont{Ahn}},
  \bibinfo{author}{\bibfnamefont{S.-C.} \bibnamefont{Kim}},
  \bibinfo{author}{\bibfnamefont{J.~I.} \bibnamefont{Wilson}},
  \bibnamefont{and} \bibinfo{author}{\bibfnamefont{P.}~\bibnamefont{John}},
  \bibinfo{journal}{The Journal of chemical physics}
  \textbf{\bibinfo{volume}{129}}, \bibinfo{pages}{234709}
  (\bibinfo{year}{2008}).

\bibitem[{sup()}]{supp_info}
\bibinfo{note}{See Supplementary Material for: (a) the bandstructures and
  Wilson loops calculated using the microscopic tight-binding model at
  different twist angles, (b) the derivations of the Chern numbers of the
  Landau levels of Dirac fermions with opposite magnetic fields, and (c) the
  details in computing the Berry phases of the high-energy bands.}

\bibitem[{\citenamefont{Trambly~de Laissardiere
  et~al.}(2010)\citenamefont{Trambly~de Laissardiere, Mayou, and
  Magaud}}]{trambly-nanoletter10}
\bibinfo{author}{\bibfnamefont{G.}~\bibnamefont{Trambly~de Laissardiere}},
  \bibinfo{author}{\bibfnamefont{D.}~\bibnamefont{Mayou}}, \bibnamefont{and}
  \bibinfo{author}{\bibfnamefont{L.}~\bibnamefont{Magaud}},
  \bibinfo{journal}{Nano letters} \textbf{\bibinfo{volume}{10}},
  \bibinfo{pages}{804} (\bibinfo{year}{2010}).

\bibitem[{\citenamefont{Moon and Koshino}(2013)}]{moon-tbg-prb13}
\bibinfo{author}{\bibfnamefont{P.}~\bibnamefont{Moon}} \bibnamefont{and}
  \bibinfo{author}{\bibfnamefont{M.}~\bibnamefont{Koshino}},
  \bibinfo{journal}{Physical Review B} \textbf{\bibinfo{volume}{87}},
  \bibinfo{pages}{205404} (\bibinfo{year}{2013}).

\bibitem[{\citenamefont{Lucignano et~al.}(2019)\citenamefont{Lucignano,
  Alf{\`e}, Cataudella, Ninno, and Cantele}}]{tbg-relaxstructure-arxiv19}
\bibinfo{author}{\bibfnamefont{P.}~\bibnamefont{Lucignano}},
  \bibinfo{author}{\bibfnamefont{D.}~\bibnamefont{Alf{\`e}}},
  \bibinfo{author}{\bibfnamefont{V.}~\bibnamefont{Cataudella}},
  \bibinfo{author}{\bibfnamefont{D.}~\bibnamefont{Ninno}}, \bibnamefont{and}
  \bibinfo{author}{\bibfnamefont{G.}~\bibnamefont{Cantele}},
  \bibinfo{journal}{arXiv preprint arXiv:1902.02690}  (\bibinfo{year}{2019}).

\bibitem[{\citenamefont{Su et~al.}(1979)\citenamefont{Su, Schrieffer, and
  Heeger}}]{ssh-1}
\bibinfo{author}{\bibfnamefont{W.~P.} \bibnamefont{Su}},
  \bibinfo{author}{\bibfnamefont{J.~R.} \bibnamefont{Schrieffer}},
  \bibnamefont{and} \bibinfo{author}{\bibfnamefont{A.~J.}
  \bibnamefont{Heeger}}, \bibinfo{journal}{Phys. Rev. Lett.}
  \textbf{\bibinfo{volume}{42}}, \bibinfo{pages}{1698} (\bibinfo{year}{1979}).

\bibitem[{\citenamefont{Vanderbilt and King-Smith}(1993)}]{vanderbilt-prb93}
\bibinfo{author}{\bibfnamefont{D.}~\bibnamefont{Vanderbilt}} \bibnamefont{and}
  \bibinfo{author}{\bibfnamefont{R.~D.} \bibnamefont{King-Smith}},
  \bibinfo{journal}{Phys. Rev. B} \textbf{\bibinfo{volume}{48}},
  \bibinfo{pages}{4442} (\bibinfo{year}{1993}).

\end{thebibliography}
\end{document}